\documentclass[prl,superscriptaddress,twocolumn,nofootinbib,10pt]{revtex4-2}

\usepackage{amsfonts,amssymb,amsmath,amsthm}
\usepackage{enumerate}
\usepackage[]{graphics,graphicx,epsfig}
\usepackage{graphicx}
\usepackage{ragged2e}
\usepackage{subcaption}
\usepackage[justification=justified,format=plain]{caption}
\usepackage{dcolumn}
\usepackage{natbib}
\usepackage{color}
\usepackage{multirow}
\usepackage{ulem}
\usepackage{bm}
\usepackage{url}
\usepackage{float}
\usepackage{mathbbol}
\usepackage{xcolor}
\usepackage{tabularx}
\usepackage{booktabs}
\usepackage{multirow}
\usepackage{gensymb}
\usepackage{mhchem} 
\usepackage{upgreek}
\usepackage{bbold}
\usepackage{xurl}
\allowdisplaybreaks[1]

\usepackage[colorlinks = true, citecolor=green, linkcolor=blue, urlcolor=blue]{hyperref}
\newcommand*{\fullref}[1]{\hyperref[{#1}]{\autoref*{#1} \nameref*{#1}}}

\usepackage{orcidlink}

\def\ket#1{\left|#1\right\rangle}

\newcommand{\norm}[1]{\left\lVert#1\right\rVert}
\newcommand{\tr}[1]{\text{Tr}\left[#1\right]}
\newcommand{\one}{\mathbb{1}}

\newtheorem{definition}{Definition}

\begin{document}


\title{Towards fault-tolerance with universal phase-error-transparent gates for high-spin cat codes}

\author{Kelvin Onggadinata \orcidlink{0000-0003-2328-8329}}
\email{kelvin.onggadinata@ntu.edu.sg}
\affiliation{School of Physical and Mathematical Sciences, Nanyang Technological University, 21 Nanyang Link, Singapore 637371, Singapore}

\author{Si Yan Koh}
\affiliation{School of Physical and Mathematical Sciences, Nanyang Technological University, 21 Nanyang Link, Singapore 637371, Singapore}

\author{Arghya Maity \orcidlink{0000-0001-6992-9473}}
\affiliation{School of Physical and Mathematical Sciences, Nanyang Technological University, 21 Nanyang Link, Singapore 637371, Singapore}

\author{Kuan Eng Johnson Goh}
\affiliation{Temasek Laboratories@Nanyang Technological University, 50 Nanyang Drive, Research Techno Plaza, BorderX Block, 9th Storey, Singapore 637553, Singapore}
\affiliation{Centre for Quantum Technologies, National University of Singapore, 3 Science Drive 2, Singapore 117543, Singapore}
\affiliation{Institute of Materials Research \& Engineering (IMRE), Agency for Science, Technology, and Research (A*STAR), 2 Fusionopolis Way, \#08-03 Innovis, Singapore 138634, Singapore}

\author{Bent Weber}
\affiliation{School of Physical and Mathematical Sciences, Nanyang Technological University, 21 Nanyang Link, Singapore 637371, Singapore}

\author{Kay Jin Lim}
\affiliation{School of Physical and Mathematical Sciences, Nanyang Technological University, 21 Nanyang Link, Singapore 637371, Singapore}

\author{Hui Khoon Ng}
\affiliation{Centre for Quantum Technologies, National University of Singapore, 3 Science Drive 2, Singapore 117543, Singapore}

\author{Teck Seng Koh
\orcidlink{0000-0002-4030-8232}}
\email{kohteckseng@ntu.edu.sg}
\affiliation{School of Physical and Mathematical Sciences, Nanyang Technological University, 21 Nanyang Link, Singapore 637371, Singapore}

\date{\today}


\begin{abstract}
High-dimensional nuclear spins offer a hardware-efficient route to quantum error correction (QEC), with the spin cat code providing intrinsic robustness against phase errors -- the dominant noise channel in donor-in-silicon architectures. 
However, realizing the full potential of this encoding requires gate operations that preserve its error-correcting properties.
In this work, we construct a universal logical gate set that is error-transparent (ET) to phase errors, and discuss its practical implementations and challenges.
The ET gates ensure that phase errors occurring stochastically during gate operations are propagated in a systematically traceable manner and remain correctable in a subsequent QEC step. 
Among the universal gate set constructed, we identify the logical $X$ gate as the primary challenge and discuss potential realization schemes. 
In addition, to fully leverage the spin cat code's advantage over an unencoded qubit, multi-tone microwave driving of the logical $CZ$ gate is essential.
Our simulations show that ET gates significantly outperform non-ET gates and may be necessary to surpass the break-even point.
We further show how logical measurement and recovery can be constructed from ET operations, and explain why state-preparation cannot be made ET.
In particular, ET measurement in the computational basis is realizable via spin parity measurement, and that error correction circuits constructed from ET operations achieve optimal error correction capacity.
Our work charts a concrete path toward full fault-tolerant quantum computation with high-dimensional nuclear spin systems.
\end{abstract}

\maketitle


\section{Introduction}\label{sec: introduction}

Quantum error correction (QEC) underpins the path to fault-tolerant quantum computation by protecting encoded quantum information from the detrimental effects of noise. 
Over the past decade, significant experimental progress has demonstrated the viability of various QEC codes, with several implementations surpassing the break-even point -- where the encoded logical qubit outlives its best unencoded physical counterpart~\cite{ofek2016extending,sivak2023real,brock2025quantum}. 
In most experimental platforms, multi-qubit codes are the dominant QEC paradigm, but their resource overhead poses serious scalability challenges~\cite{steane2003overhead, diVincenzo2009fault, takita2017experimental, chao2018fault}.
A hardware-efficient alternative is to encode a logical qubit within a single higher-dimensional system, eliminating the overhead of multi-qubit encoding at the physical level. 
Encodings in high-dimensional spin systems are the finite-dimensional analogues of bosonic codes~\cite{cochrane1999macroscopically,gottesman2001encoding,grimsmo2020quantum}, where a logical qubit is encoded in an infinite-dimensional oscillator mode, and many theoretical tools developed in that context can be systematically adapted to the high-spin case.

Recent experiments with a spin-7/2 antimony ($\ce{^{123}Sb}$) nucleus in silicon have demonstrated high-fidelity generation of Schr\"{o}dinger cat states~\cite{yu2025Schrodinger} as well as coherent control of its 16-dimensional Hilbert space when coupled to an electron ancilla~\cite{fuentes2024navigating}.
Electron spin resonance spectroscopy of a spin-9/2 germanium ($\ce{^{73}Ge}$) nucleus via a multi-electron quantum-dot ancilla has also been demonstrated~\cite{steinacker2026coupling}.
Combined with theoretical proposals for high-dimensional spin encodings in donors~\cite{gross2021designing, gross2024hardware}, molecular spins~\cite{chiesa2020molecular,mezzadri2024fault} and neutral atoms~\cite{omanakuttan2024fault, debry2025error, kusano2026spin}, these recent developments signal a significant step towards hardware-efficient QEC with high-dimensional spins.

In this work, we focus on the donor-in-silicon platform, in which group-V donor atoms possess a loosely-bound electron that can serve as a functional ancilla or be ionized to mitigate decoherence when not required.
These donor nuclear spins are remarkably long-lived, with lifetimes of several hours for phosphorus~\cite{muhonen2014storing}. 
Coherence times for phosphorus donors can reach up to $T_{2DD} \approx 3$~minutes with dynamical decoupling~\cite{pla2013high, morello2020donor}, while recent measurements on high-spin antimony donors yielded $T_2^* \approx 20-50$~ms~\cite{fuentes2024navigating, yu2025Schrodinger}, making dephasing the dominant source of error.
Given this intrinsically biased noise environment, the spin cat code, with its inherent robustness against phase errors~\cite{chiesa2020molecular,gross2024hardware}, is particularly well-suited for donor-based QEC.

While QEC schemes protect stored quantum states from noise, errors arising during gate operations can accumulate and propagate through a circuit, compromising the reliability of quantum information processing. 
Achieving fully fault-tolerant architectures remains an open challenge, requiring that every component, from gate operations to state preparation, measurement and error correction, individually satisfy stringent fault-tolerance conditions.
A fault-tolerant gate must not propagate errors in an uncontrollable manner beyond what the code can subsequently correct~\cite{gottesman2009introduction}.
Existing studies on high-spin codes have constructed explicit pulse sequences for encoding, decoding and correction, but without fault-tolerant gadgets for these operations~\cite{lim2023fault}, or have addressed fault-tolerant gate implementations in neutral atoms~\cite{omanakuttan2024fault} rather than donor systems.
The fault-tolerance potential of the spin cat encoding in donor systems thus remains largely unexplored.

For high-spin codes, certain conventional fault-tolerance formulations, such as transversality, do not apply here due to the lack of multi-component physical structure.
Rather than demanding a fully fault-tolerant formulation from the outset, practical schemes that demonstrate key ideas towards this goal have been proposed, drawing in particular from notions of fault-tolerance developed in the bosonic QEC literature, including error- or bias-preserving~\cite{puri2020bias} and path independent~\cite{ma2020path} operations.
Of particular significance is the concept of error-transparent (ET)  operations~\cite{vy2013error, kapit2018error}, in which operations are specially designed such that errors can propagate in a controlled and correctable manner, preserving the error structure needed for subsequent QEC.

In this work, we construct a universal logical gate set satisfying the ET criterion and discuss its practical implementation in the donor-in-silicon platform.
Our results and proposals can be easily generalized to other high-dimensional quantum computing platforms with biased noise.
From numerical simulations of error correction, we find that ET gates significantly outperform non-ET gates and may be necessary to surpass the break-even point. 
We further show how logical measurement and error corrections can be constructed from ET operations, and identify logical state preparation as the operation that requires non-ET gates. 
The ET measurement in the computational basis can be realized with spin parity measurements, and that error correction circuits constructed from ET operations achieve optimal error correction capacity. 
Taken together, our results delineate a concrete and realistic path towards full fault-tolerance within this hardware-efficient framework.


\section{Results}\label{sec: preliminaries}

\subsection{High-spin donor system}\label{subsec: donor system}

Donors implanted in silicon belong to group-V elements, which have a single, loosely-bound valence electron. 
Stable elements in this group that have a nuclear spin $I>1/2$ are arsenic ($\ce{^{33}As}$, $I=3/2$), the isotopes of antimony -- antimony-121 ($\ce{^{121}Sb}$, $I=5/2$) and antimony-123 ($\ce{^{123}Sb}$, $I=7/2$) -- and bismuth ($\ce{^{83}Bi}$, $I=9/2$).
The Hamiltonian for a neutral donor $\hat{H}_{D^0}$ comprises the ionized donor term $\hat{H}_{D^+}$ and the term $\hat{H}_\text{en}$ accounting for the valence electron and its hyperfine interaction with the nucleus:
\begin{equation}
\hat{H}_{D^0} = \hat{H}_{D^+} + \hat{H}_\text{en}\, ,
\end{equation}
where 
\begin{align}
\hat{H}_{D^+} &= - \gamma_n\hat{I}_z B_0 + \sum_{\alpha,\beta\in\{x,y,z\}}Q_{\alpha\beta}\hat{I}_\alpha\hat{I}_\beta \, \\
\hat{H}_\text{en} &= \gamma_e\hat{S}_z B_0 + A\hat{\mathbf{S}} \cdot \hat{\mathbf{I}} \, .
\end{align}
Here, a longitudinal magnetic field $\vec{B}= B_0\hat z$ produces Zeeman-split levels according to the nuclear (electron) gyromagnetic ratio $\gamma_n$ ($\gamma_e$), $\hat{\mathbf{I}} = [\hat{I}_x,\hat{I}_y,\hat{I}_z]$ ($\hat{\mathbf{S}} = [\hat{S}_x,\hat{S}_y,\hat{S}_z]$) is the nuclear (electron) spin operator in the Cartesian axes, and $A$ is the hyperfine interaction strength. 
The non-spherical charge distribution of nuclei with spin $I\geq 3/2$, characterized by the quadrupole moment $q_n$, couples to an electric field gradient $\mathcal{V}_{\alpha\beta}$, giving rise to the quadrupole energy $Q_{\alpha\beta} = \frac{eq_n\mathcal{V}_{\alpha\beta}}{2I(I-1)h}$.
This interaction introduces unequal shifts to the Zeeman-split levels, enabling individual addressability of the nuclear spin states. 
In Table \ref{tab: group V donor parameters} [see Supplemental Material (SM) Sec. A], we provide the values of the physical parameters associated with the different group-V donors.

\begin{figure}
\centering
\includegraphics[width=1\linewidth]{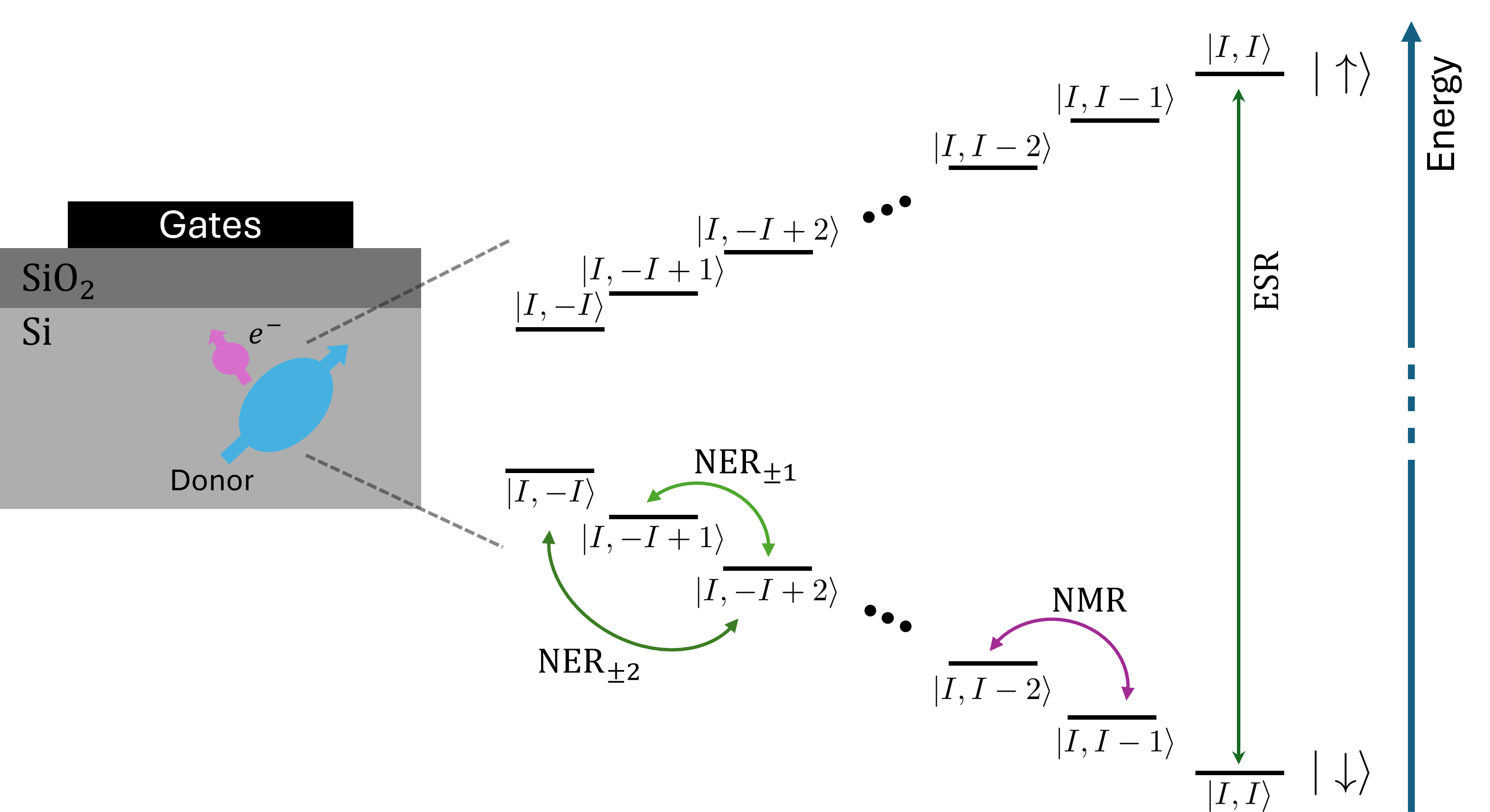}
\caption{\justifying{Diagram of the donor and electron system in silicon and their energy levels. The eigenstates $|s\rangle \otimes |I,m_I\rangle$ are split into two branches with electron spin $s=\downarrow$ ($s=\uparrow$) corresponding to the lower (upper) branch. Electron spin resonance (ESR) controls inter-branch transitions, while transitions between nuclear spin levels within each branch can be controlled with resonant magnetic fields (NMR). The quadrupole interaction enables transitions via electric fields on resonance with nuclear spacings of $\Delta m_I = \pm1$ and $\Delta m_I =\pm2$ ($\mathrm{NER}_{\pm1}, \mathrm{NER}_{\pm2}$).} 
}
\label{fig: system and energy diagram}
\end{figure}

Under a strong magnetic field $B_0 \sim 1$ T, the interaction terms follow the scaling $\gamma_eB_0 \gg \gamma_n B_0 \gg A \gg Q_{\alpha\beta}$. 
As such, the system eigenstates are well approximated by the tensor product of the electron spin state $\{|\downarrow\rangle,|\uparrow\rangle\}$ and the nuclear spin state $\{|I,m_I\rangle\}$ in the longitudinal or $z$-basis, with $m_I=-I,-I+1,\dots, I-1,I$ (see Fig. \ref{fig: system and energy diagram}). 
The total dimension of the system is $d = 2d_n$, where $d_n = 2I+1$ is the dimension of the nuclear spin space in which we encode the logical qubit.
The quadrupole interaction lifts the degeneracy of nuclear spin transitions, allowing for individual addressability of nuclear spin states in the absence of the valence electron in ionized donors. 
The electron can then be independently leveraged as an ancilla for quantum nondemolition (QND) measurement of the nuclear spin, as well as error detection and correction.

Manipulation of the state is achieved by driving a transversal a.c. electric or magnetic field with Hamiltonian of the form
\begin{equation}\label{eq: control hamiltonian}
\hat{H}_{\text{d}} = (-\gamma_n\hat{I}_x + \gamma_e\hat{S}_x)\sum_{k}B_{d,k}(t)\cos(\omega_k t + \phi_k)\,, 
\end{equation}
where multi-frequency tone of amplitude $B_{d,k}(t)$ is applied with angular frequency $\omega_k$ and phase $\phi_k$.
Electron spin resonance (ESR) is achieved by a transverse magnetic field drive at one of the transition frequencies $\omega_{m_I}^{\text{ESR}}$, each conditional on the nuclear spin projection $m_I$, to flip the electron state $|\downarrow\rangle \leftrightarrow |\uparrow\rangle$. 
Nuclear spin transitions between neighbouring eigenstates ($\Delta m_I = \pm 1$) are driven by nuclear magnetic resonance (NMR), via an oscillating magnetic field coupling to the off-diagonal matrix elements of $\hat{I}_x$.
An oscillating electric field provides an alternative control channel through nuclear electric resonance (NER), which exploits modulation of the electric quadrupole interaction rather than the nuclear Zeeman coupling. 
This activates transitions via $\hat{I}_z\hat{I}_\pm$ for $\Delta m_I = \pm 1$ ($\text{NER}_{\pm 1}$), and via $\hat{I}^2_\pm$ for $\Delta m_I = \pm 2$ ($\text{NER}_{\pm 2}$), with the microscopic mechanism arising from a time-dependent electric-field gradient at the nucleus.


\subsection{Error model}\label{subsec: error model}

Dephasing of donor spins in silicon arises from magnetic and electric noise.
The former is primarily due to spectral diffusion from the fluctuating nuclear spin bath of $\ce{^{29}Si}$ in natural silicon, which can be mitigated by isotopic purification.
Electric noise couples to the donor through two channels.
For the ionized donor in which the logical states are encoded, charge noise modulates the electric field gradient at the nucleus, shifting the quadrupole energies; for the neutral donor, charge noise modulates the electron wavefunction and causes dephasing via the hyperfine interaction.

The error model for dephasing in high-spin donor systems is justified by the hierarchy of energy scales in the system Hamiltonian.
The dominant contribution is the nuclear Zeeman term $\gamma_n B_0 \hat{I}_z$, while the quadrupole interaction, comprising terms quadratic in the spin operators $\hat{I}_\alpha \hat{I}_\beta$, enters as a perturbative correction 
satisfying $\gamma_n B_0 \gg Q_{\alpha\beta}$.
Stochastic fluctuations therefore couple predominantly through $\hat{I}_z$, justifying truncation of the error set to powers of $\hat{I}_z$.

Dephasing errors are thus modeled by the spin operator $\hat{I}_z$, and we denote the set of rank-$l$ dephasing errors as $\mathcal{E}_z^{[l]} \equiv \{\one,\hat{I}_z,\hat{I}_z^2,\dots \hat{I}_z^l\}$, $l \in \mathbb{N}$.
It has been shown that encoding into spin coherent states perpendicular to the $z$-axis offer maximal protection against such dephasing~\cite{chiesa2020molecular,gross2024hardware}.


\subsection{High-spin encoding}\label{subsec: spin cat code}

Spin coherent states $|I,\pm I\rangle_{\hat{\mathbf{v}}}$ are extremal eigenstates of the spin operator $\hat{I}_{\hat{\mathbf{v}}} \equiv v_x\hat{I}_x + v_y\hat{I}_y + v_z\hat{I}_z$, $\hat{\mathbf{v}} = [v_x,v_y,v_z] \in \mathbb{R}^3$, $\norm{\hat{\mathbf{v}}} = 1$, such that $\hat{I}_{\hat{\mathbf{v}}}|I,\pm I\rangle_{\hat{\mathbf{v}}} = \pm I |I,\pm I\rangle_{\hat{\mathbf{v}}}$, and has been regarded as the spin correspondence to bosonic coherent states \cite{radcliffe1971properties, arecchi1972atomic}.
In analogy to the Schr\"{o}dinger cat codes from bosonic QEC, we define the logical codewords as superpositions of the two spin coherent states along the $x$-axis, which we refer to as the {\it spin cat-$x$ code}:
\begin{equation}
|\mu_L\rangle \equiv \frac{1}{\sqrt{2}}\Big[|I,I\rangle_x + (-1)^\mu |I,-I\rangle_x\Big], \quad \mu=0,1\, .
\end{equation}
The spin cat-$x$ code satisfies the Knill-Laflamme condition~\cite{knill1997theory} for rank-$r$ dephasing errors with the set $\mathcal{E}^{[r]}_z$, where $r = \lfloor I\rfloor = d_n/2-1$. 
In the SM Sec. B, we discuss further the equivalence of the spin cat-$x$ code with the spin binomial code and its relations with rotationally symmetric spin codes, in analogy to the bosonic case~\cite{grimsmo2020quantum}.

\begin{figure*}
\centering
\begin{subfigure}{0.35\linewidth}
\centering
  \includegraphics[width=0.95\linewidth]{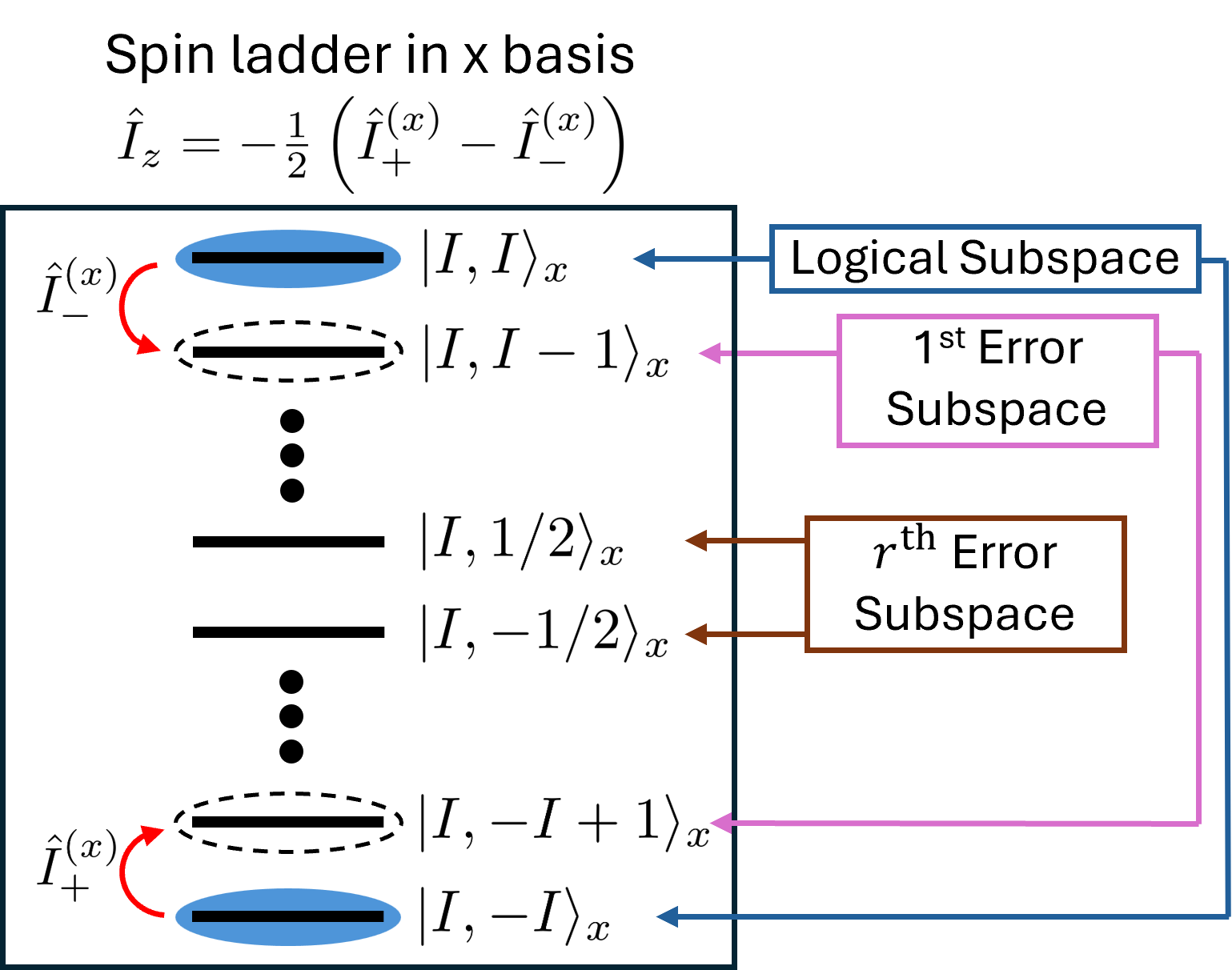}
  \caption{}
  \label{fig: ET concept error ladder}
\end{subfigure}
\begin{subfigure}{0.60\linewidth}
\centering
  \includegraphics[width=0.95\linewidth]{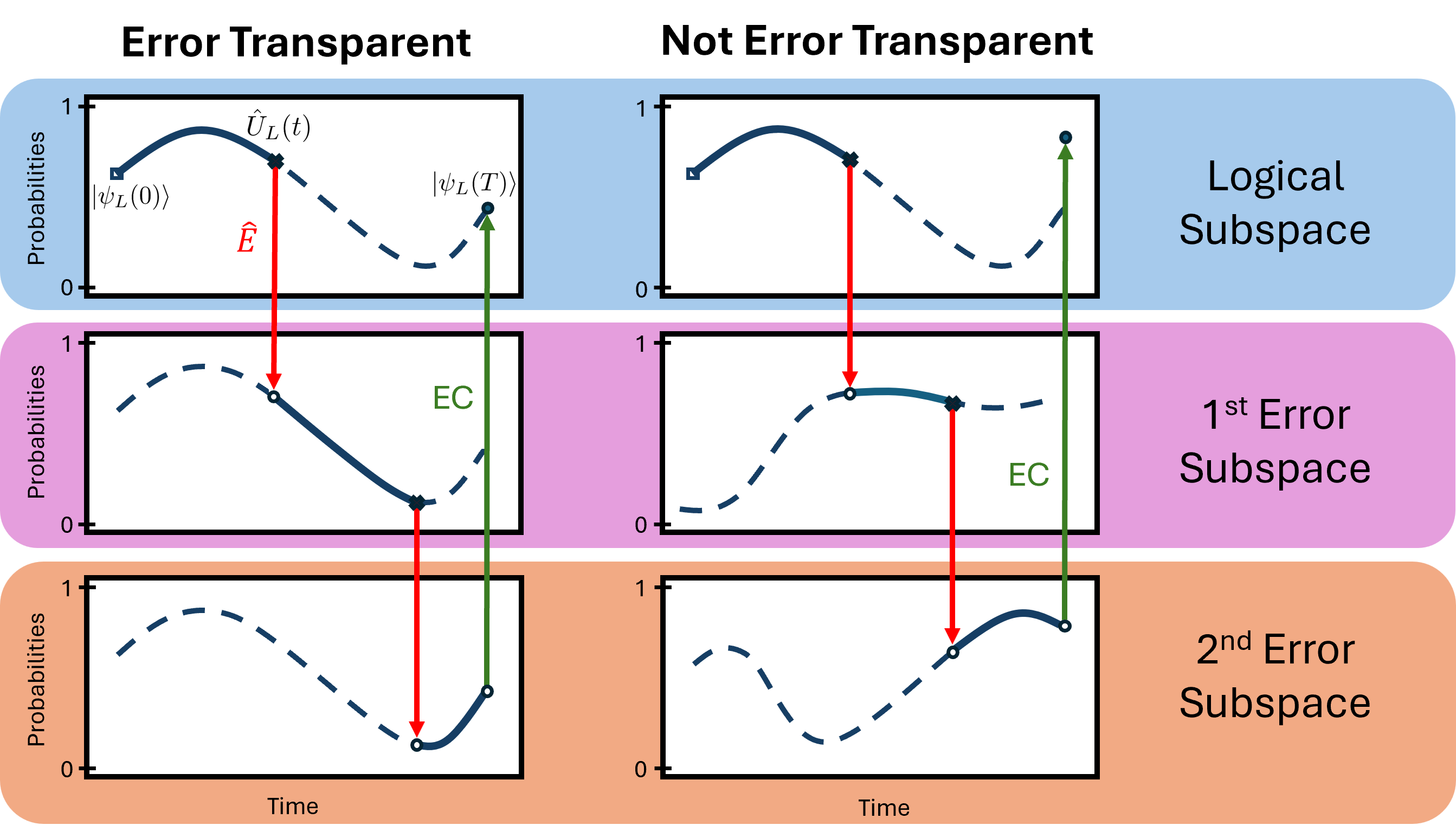}
  \caption{}
  \label{fig: ET and not ET diagram} 
\end{subfigure}
\caption{\justifying{(a) Dephasing errors act as spin ladder operations in the $\{|I,m_I\rangle_x\}$ basis, with each $|m_I|$ forming the support state for the logical subspace and error subspaces. (b) Conceptual illustration of error-transparent (ET) and non-ET operations. The red arrow illustrates the action of an error operator causing jumps from error subspaces of lower rank to higher ones. ET operations preserve the intended trajectories/evolutions in the logical subspace in each of the error subspaces. Performing error correction (represented by green arrow) at any time then brings the transformed state in the error subspaces to the appropriate state in the logical subspace.}}
\label{fig: ET concept diagram}
\end{figure*}

The error-correcting structure becomes clearer upon expressing $\hat{I}_z$ in the $x$-basis, $\hat{I}_z = -\frac{1}{2}\left(\hat{I}_+^{(x)} - \hat{I}^{(x)}_-\right)$, which shows that a single phase error shifts each codeword by one step into an orthogonal subspace: 
$|I,\pm I\rangle_x \to |I,\pm(I-1)\rangle_x$, leaving the logical information in a detectable but correctable error subspace.
Since the dimension of the nuclear spin $d_n$ limits the number of shifts possible, the codewords remain correctable up to $r$ errors (see Fig. \ref{fig: ET concept error ladder}).
The correctable rank $r$ is equivalent to the {\it weight} in conventional multi-qubit QEC, and the {\it code distance} $p = 2r+1$~\cite{nielsen2010quantum}. 

The concept of error subspaces provides a systematic framework for tracking how error operators map the codewords, and underpins both the error correction protocol and the error-transparency criterion discussed in the next section.
The logical and error subspaces are spanned by the image of the codewords under the error operator of orders up to rank-$r$, that is, $\{\hat{I}_z^k|0_L\rangle,\hat{I}_z^k|1_L\rangle\}_{k=0}^r$.
Since these states are in general not orthonormal, we perform a Gram-Schmidt procedure and denote the resulting basis by $\{|\mathbb{E}_{\mu}^k\rangle\}_{k=0}^r$, where
\begin{equation}
|\mathbb{E}_{\mu}^k\rangle = \frac{(-1)^k}{\sqrt{2}}\Big[|I,I-k\rangle_x + (-1)^\mu|I,-I+k\rangle_x\Big]\, ,\quad \mu=0,1\, .
\end{equation}
These states are spin cat-$x$ states of decreasing magnitude $|m_I|=|I-k|$ with increasing error of order $k$ (see Fig. \ref{fig: ET concept error ladder}). 
We refer to these states as {\it errorwords} when $k > 0$, with $|\mathbb{E}_\mu^0\rangle = |\mu_L\rangle$ being the original codewords.
Hence, we denote the $k$-th error subspace as $\mathcal{P}_k = \text{span}\{|\mathbb{E}_{0}^k\rangle,|\mathbb{E}_1^k\rangle\}$, with $\mathcal{P}_0 = \mathcal{P}_\mathcal{C}$ being the logical subspace (codespace).
The projective measurement $\hat{P}_k = |\mathbb{E}_0^k\rangle\langle \mathbb{E}_0^k| + |\mathbb{E}_1^k\rangle\langle \mathbb{E}_1^k|$
thus determines the $k$-th error subspace population. 
Table \ref{tab: spin cat-$x$ error basis} (SM Sec. B) shows the errorwords in the $z$-basis explicitly.


\subsection{Universal error-transparent gates}\label{sec: universal ET gates}

\subsubsection{Error transparency criterion}\label{subsec: ET criteria}

ET gates were first proposed in Refs.~\cite{vy2013error, kapit2018error} and further explored in the field of bosonic QEC codes~\cite{ma2020error,wetherbee2025mathematical}. 
Consider a unitary gate $\hat{U}(t_f,t_i)$ associated with a generating Hamiltonian $\hat{H}(t)$, and an error that could occur in the time interval $[t_i, t_f]$, acting on a state $\ket{\psi_L}$ in the logical subspace.
If the Hamiltonian were engineered such that evolution trajectories within all error subspaces were identical to the  trajectory in the logical subspace, then the error state can be corrected back to the logical state at the final instant (see Fig. \ref{fig: ET and not ET diagram}).
As such, it does not matter at what point in the time interval the error occurred.
The error transparency criterion is thus
\begin{equation}
    [\hat{E}_j,\hat{H}(t)]|\psi_L\rangle = 0\, ,\quad \forall \, |\psi_L\rangle, t \in [t_i,t_f], \hat{E}_j\in\mathcal{E}\, ,
\end{equation}
for a given error set $\mathcal{E}$.
Due to the commutativity of the generating Hamiltonian with the error $\hat{E}_j$, the effect of the error occurring at a time $t$ within the interval $[t_i, t_f]$ is the same as the error occurring at the final instant: $\hat{U}(t_f,t)\hat{E}_j \hat{U}(t,t_i)|\psi_L\rangle = \hat{E}_j\hat{U}(t_f,t_i)|\psi_L\rangle$, as if the gate were transparent to the error, hence the nomenclature.
This ensures that any error occurring within $[t_i, t_f]$ remains correctable by a subsequent QEC step, thus satisfying a necessary condition of fault-tolerant operation. 
Note that there exists a more general ET criterion~\cite{ma2020error}.
However, in this work the ET operations we propose satisfy the above formulation.

\subsubsection{Universal phase-error-transparent gate set}\label{subsec: universal ET gate set}

We now show the construction of a set of universal gates for the spin cat-$x$ code that are phase-error-transparent against rank-$r$ dephasing errors $\mathcal{E}_z^{[r]}$. 
The universal logical gate set $\{\overline{Z}(\theta), \overline{CZ}, \overline{X}(\phi)\}$ contains the logical phase gate with arbitrary rotation angle $\theta$, the logical controlled-$Z$ gate, and the logical $X$ rotation gate with angle $\phi$. 
Here, an overline indicates a logical gate.
Gates with arbitrary rotation around two axes can generate an arbitrary single-qubit gate, and adding any entangling two-qubit gate is sufficient for universality. 
Thus, the above set is overcomplete; it is sufficient to have $\overline{X}(\frac{\pi}{2})$ -- the logical $\sqrt{NOT}$ gate -- instead of an arbitrary rotation to achieve universality~\cite{mckay2017efficient}.
The $\overline{Z}(\theta)$ and $\overline{CZ}$ gates have a straightforward ET construction by taking advantage of the symmetry of the spin cat-$x$ code and its errorwords.
On the other hand, the $\overline{X}(\phi)$ gate, which constitutes an amplitude-mixing operation, does not have a trivial construction.

\subsubsection{Error-transparent $\overline{Z}$ gate}

The logical phase gate $\overline{Z}(\theta)$ is defined by the following transformation: $\overline{Z}(\theta)|\mu_L\rangle = e^{i\theta\delta_{\mu,1}}|\mu_L\rangle$, where $\delta_{\mu,1}$ is the Kronecker delta.
Its realization can be achieved via the so-called {\it virtual SNAP} (selective number-dependent arbitrary phase) operation~\cite{yu2025Schrodinger}, which is a diagonal unitary with an arbitrary phase associated with each nuclear spin state in the $z$-basis:
\begin{equation}
S(\bm{\zeta}) = \sum_{m_I=-I}^I e^{i\zeta_{m_I}} |I,m_I\rangle \langle I,m_I|\, ,
\end{equation}
where $\bm{\zeta} = (\zeta_{-I},\zeta_{-I+1},\dots,\zeta_{I})$.
This operation is realized virtually; an appropriate phase update is done on a software-defined generalized rotating frame (GRF).
Since the support states (in the $z$-basis) of the state $|\mathbb{E}_1^k\rangle$ of each error subspace are the same as those of the logical state $|1_L\rangle$, i.e. $\text{supp}_{m_I}(|\mathbb{E}_1^k\rangle) = \text{supp}_{m_I}(|1_L\rangle)$ for all $k=1,\dots,r$, by applying identical phases on the support states of $|\mathbb{E}_1^k\rangle$ for all $k$, the SNAP gate becomes the ET logical phase gate,
\begin{equation}
\overline{Z}(\theta) = S(0,\theta,0,\theta,\dots,0,\theta).
\end{equation}
This is efficient to implement classically since the computational cost scales linearly with the number of states involved.

\subsubsection{Error-transparent $\overline{CZ}$ gate}\label{subsec: ET $CZ$ gate}

For two-qubit operations, we consider the two donor nuclei, labelled as $\mathrm{n}_1$ and $\mathrm{n}_2$, with a shared electron such that the hyperfine coupling with $\mathrm{n}_1$ is much stronger than that with $\mathrm{n}_2$.
This can be achieved experimentally by displacing the electron wavefunction with gate-applied electric fields so that most of the wavefunction overlaps with one of the donor nuclei~\cite{madzik2022precision}.
The interaction between the nuclei is mediated by the exchange interaction of each nucleus with the electron. 
By exploiting this interaction, Ref.~\cite{madzik2022precision} reported a $CZ$ gate between two nuclear spins of $\ce{^{31}P}$ donors. 
Since all electron-nuclei transitions are individually addressable, an ESR $2\pi$-pulse can be applied such that it is conditional on both nuclei being spin-down.
As a result, the electron acquires a phase $e^{i\pi}$ on the selected spin-down state of two nuclear spins. 
The effective transformation, upon tracing out the electron, is equivalent to a phase flip on two-qubit spin-down state, leaving other two-qubit states unchanged, thus achieving a $CZ$ gate.

Building on the same ESR-based construction, an ET logical $\overline{CZ}$ gate can be realized for the spin cat-$x$ code. 
The $\overline{CZ}$ gate acts on the logical codewords as $\overline{CZ}|\mu_L,\nu_L\rangle =  (-1)^{\delta_{\mu,1}\delta_{\nu,1}}|\mu_L,\nu_L\rangle$, implementing a conditional phase flip exclusively on the $|1_L,1_L\rangle$ component. 
At its core, this is realized by applying ESR $2\pi$-pulses pulses conditioned on each support state of $|1_L,1_L\rangle$.
Since $|1_L,1_L\rangle$ and $|\mathbb{E}_1^k,\mathbb{E}_1^k\rangle$ share the same support states, and they are disjoint from $\text{supp}_{m_I}(|\mathbb{E}^k_\mu,\mathbb{E}_\nu^l\rangle)$, $\mu,\nu\neq 1$ for all $k,l$, the construction is ET by design.

We illustrate this for two logical qubits encoded in $\ce{^{75}As}$ ($I = 3/2$) donors 
(see Fig.~3). The support of $|1_L,1_L\rangle$, expressed in the $\hat{I}_z$ eigenbasis, 
consists of four states:
$\text{supp}_{m_I}(|1_L,1_L\rangle) = \{|-\tfrac{1}{2};-\tfrac{1}{2}\rangle, 
|-\tfrac{1}{2};\tfrac{3}{2}\rangle, |\tfrac{3}{2};-\tfrac{1}{2}\rangle, 
|\tfrac{3}{2};\tfrac{3}{2}\rangle\}$.
Accordingly, four ESR pulses are required. 
For higher spin donors, the construction generalises 
straightforwardly, with the number of required ESR pulses scaling as 
$(I+\tfrac{1}{2})^2$. Assuming a driving field of magnitude $B_d = 0.1$ mT, each ESR $2\pi$-pulse takes 
approximately $0.7~\mu$s, well within the electron coherence time~\cite{muhonen2014storing}. Even though Fig.~3 illustrates sequential pulses, multi-tone driving enables efficient realization in a single step, thus minimizing gate time. We note, however, that this operation is not ET with respect to charge noise on the electron, a point we return to in the Discussion.

\begin{figure}
\centering
\includegraphics[width=1\linewidth]{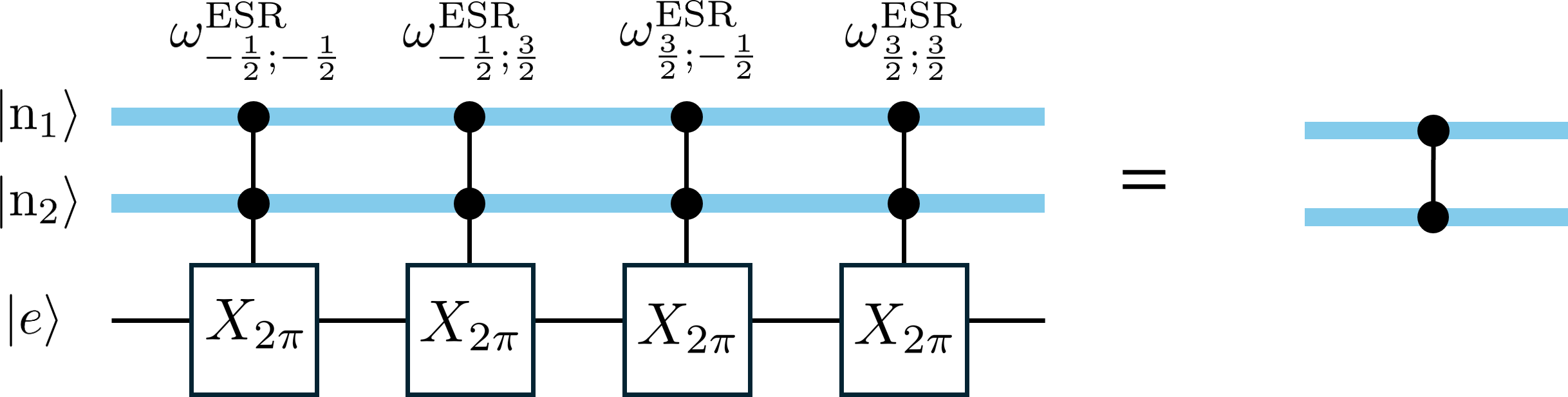}
\caption{\justifying{Error-transparent implementation of logical $CZ$ gate for $I=3/2$. A sequence of four consecutive ESR-based $2\pi$ rotations applied on the electron ancilla produces a geometric phase conditional on the support states of the $|1_L\rangle$ state, resulting in an ET $\overline{CZ}$ gate. Each thick blue line corresponds to a high-spin donor ($d$-level system) with state labels $|\mathrm{n}_{k}\rangle$, $k=1,2$, while the thin black line corresponds to the electron system with state label $|e\rangle$. The ESR frequencies are denoted by $\omega^{\text{ESR}}_{m_I,m_I'}$ with $m_I,m_I'$ being the conditional spin states on $\mathrm{n}_1,\mathrm{n}_2$, respectively.}
}
\label{fig: ET CZ d4 protocol}
\end{figure}

\subsubsection{Error-transparent $\overline{X}$ gate}\label{subsec: ET X gate}

The phase gates constructed above are relatively straightforward to make 
error-transparent, as they act exclusively on the support of $|\mathbb{E}_1^k\rangle$, which is disjoint from the support of $|\mathbb{E}_0^k\rangle$.
This disjointness makes it straightforward to enforce the ET condition by restricting the operation to the relevant support.
The logical $\overline{X}$ gate, however, presents a fundamentally greater challenge: it mixes amplitudes between $|0_L\rangle$ and $|1_L\rangle$, and can therefore not be constructed by acting on the support of a single codeword alone.

Nevertheless, an ET $\overline{X}$ gate can be constructed by extending the action of an $\overline{X}$ gate on the codewords to the errorwords, such that the Hamiltonian
\begin{equation}
\hat{H}_{\overline{X}} = \frac{\Omega}{2}\sum_{k=0}^r |\mathbb{E}_0^k\rangle \langle \mathbb{E}_1^k| +  |\mathbb{E}_1^k\rangle \langle \mathbb{E}_0^k|\, ,
\end{equation}
generates the ET unitary.
Since the errorwords are orthogonal, such a construction is ET by design.
The logical gate is then $\overline{X}(\phi) = \exp(-it\hat{H}_{\overline{X}}) = \exp(-i\phi \hat{H}_{\overline{X}}/\Omega)$, where $\phi = \Omega t$, where $\Omega$ is the Rabi frequency. 
By decomposing the above in the $x$-basis, we obtain a more instructive expression as linear combinations of odd-powers of $\hat{I}_x$:
\begin{eqnarray}
& & \hat{H}_{\overline{X}}/\Omega \nonumber \\ &=&  \frac{1}{2}\sum_{k=0}^r |I,I-k\rangle_x\langle I,I-k| - |I,-I+k\rangle_x\langle I,-I+k| \nonumber \\
&=& \sum_{\substack{n=1\\n\in \text{ odd}}}^{2I}c^{(I)}_{n}\hat{I}_x^n\, , \label{eq: ET X Hamiltonian linear decomposition}
\end{eqnarray}

where $c_{n}^{(I)}$ are real coefficients whose solution can be found from a system of $\lfloor I \rfloor +1$ linear equations. 
We show the derivation and explicit decomposition for all four high-spin cases $\frac{3}{2} \leq I \leq \frac{9}{2}$ in Methods.

\begin{figure*}
\centering
\begin{subfigure}{0.49\textwidth}
\centering
  \includegraphics[width=0.99\linewidth]{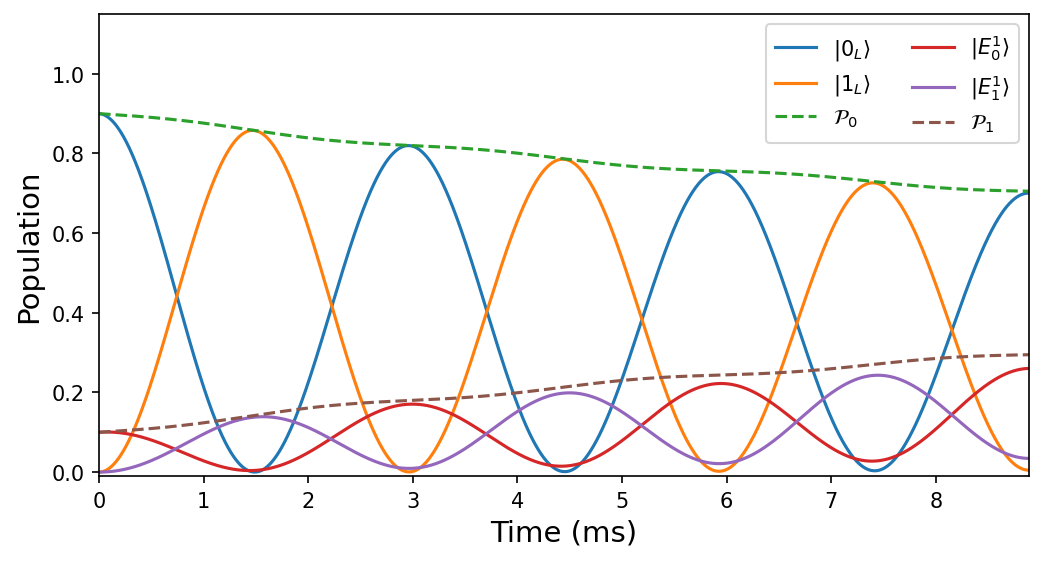}
  \caption{}
  \label{fig: ET X gate rotation illustrate}
\end{subfigure}
\begin{subfigure}{0.5\textwidth}
\centering
  \includegraphics[width=0.99\linewidth]{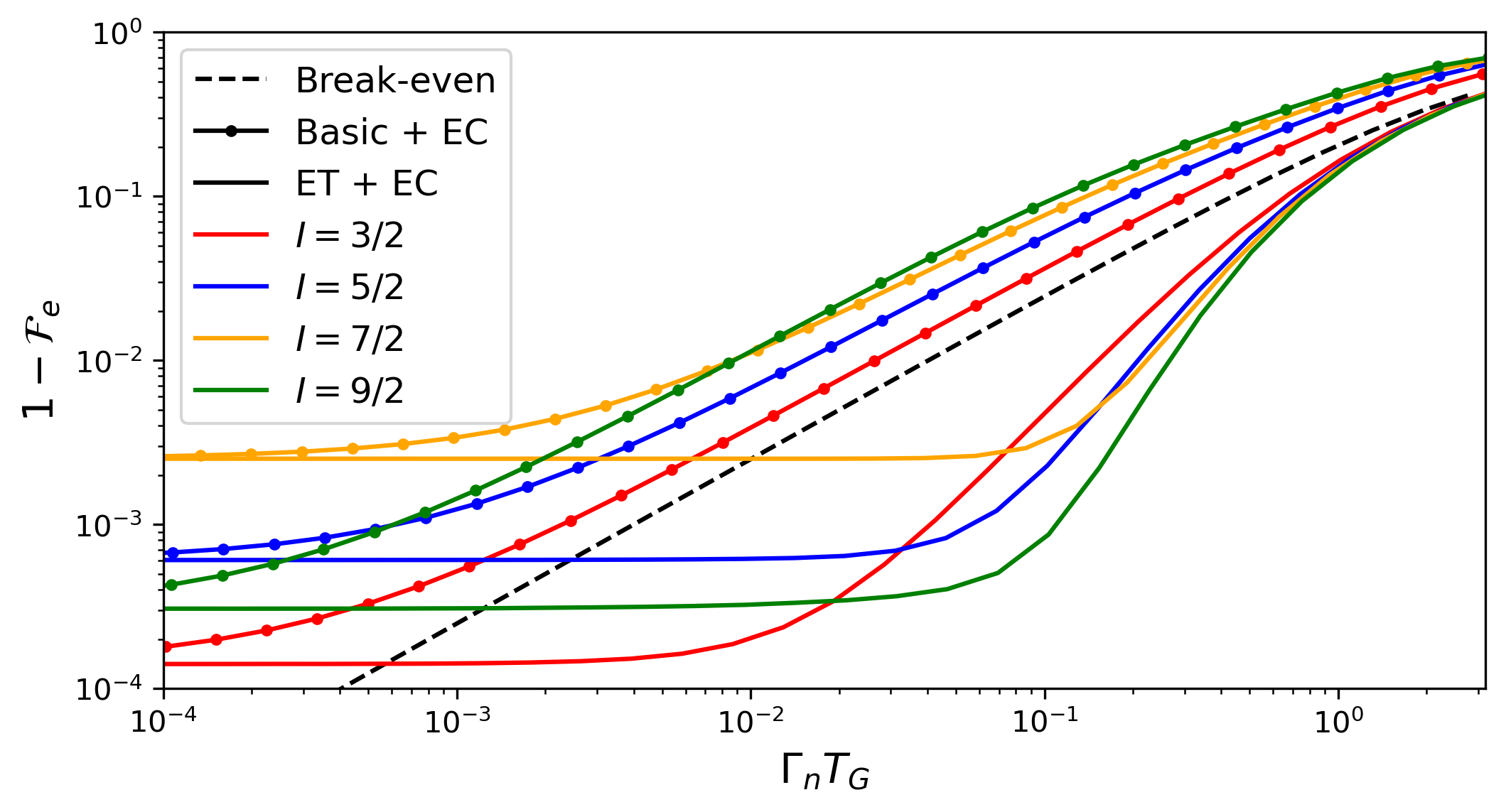}
  \caption{}
  \label{fig: ET X gate fidelities}
\end{subfigure}
\begin{subfigure}{1\textwidth}
\centering
  \includegraphics[width=0.99\linewidth]{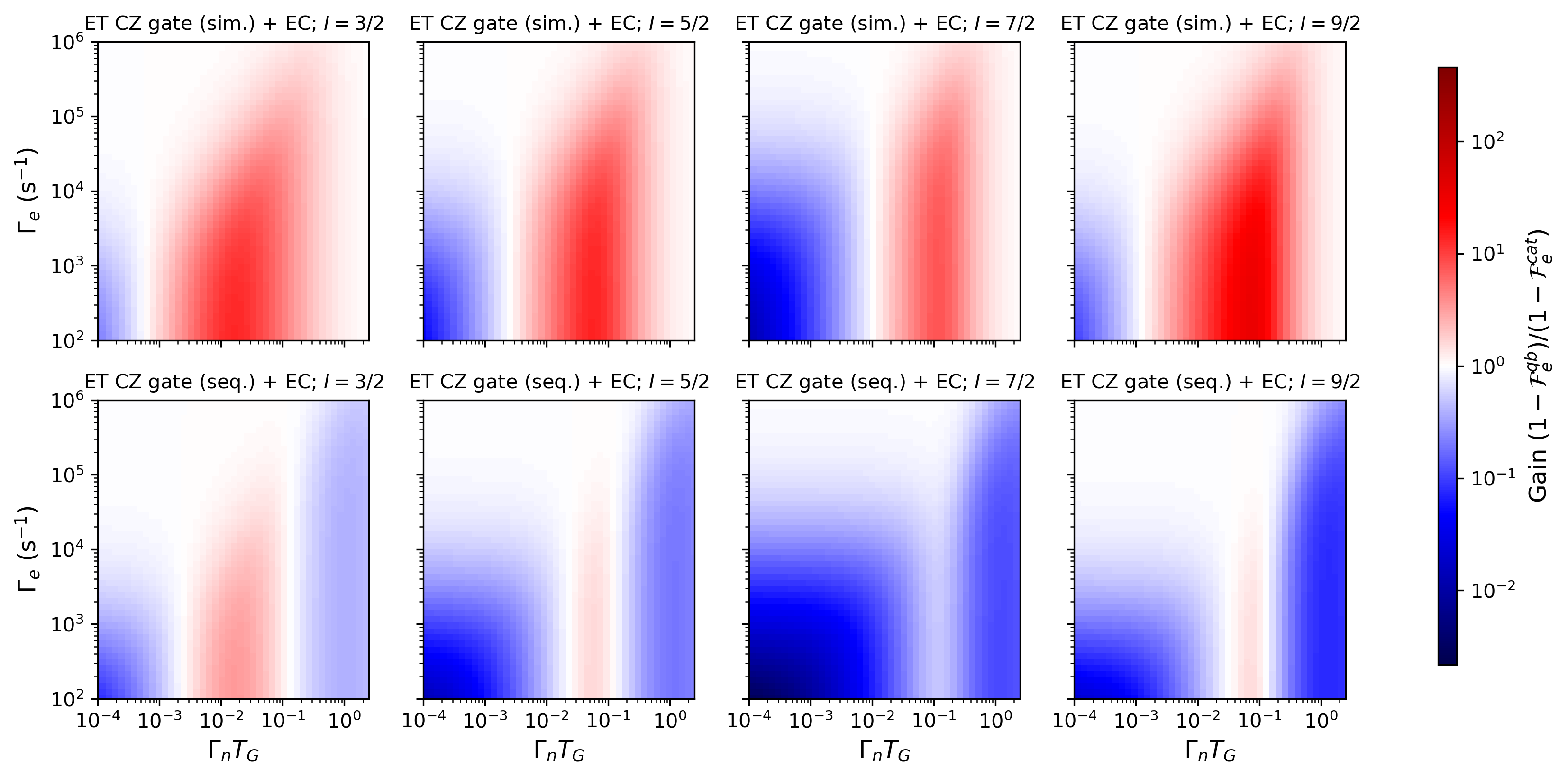}
  \caption{}
  \label{fig: ET $CZ$ gate performance complete}
\end{subfigure}
\caption{\justifying{Evolution and numerical performances of ET gates. (a) State evolution under an ET $\overline{X}$ gate rotation with initial state $|\psi_0\rangle = \sqrt{0.9}|0_L\rangle + \sqrt{0.1}|\mathbb{E}_0^1\rangle$. The error-transparent evolution is demonstrated through the synchronized population oscillations in the respective logical and first error subspace. The dashed lines indicate the total population of the logical subspace $\mathcal{P}_0$ and the first error subspace $\mathcal{P}_1$. (b) Numerical performance of basic (solid line with points) and ET (solid line) $\overline{X}_{\pi/2}$ gate followed by noisy error correction for different spin $\frac{3}{2} \leq I \leq \frac{9}{2}$ (each spin is color coded). The performance is evaluated using entanglement infidelities $1-\mathcal{F}_e$ as a function of nuclear dephasing noise strength $\Gamma_n T_G$. The break-even (dashed) line corresponds to the fidelities of the uncorrected spin-1/2 system. Points below the break-even line correspond to the regions where encoding and error correction advantage is observed. (c) ET $CZ$ gate performance for spins $\frac{3}{2} \leq I \leq \frac{9}{2}$ with ESR pulses implemented simultaneously (top) and sequentially (bottom). We evaluate the performance via gain $G = (1-\mathcal{F}_e^{\text{qb}})/(1-\mathcal{F}_e^{\text{cat}})$ defined as the ratio of infidelities between the uncorrected spin-1/2 system and spin cat-x code. The red region indicates $G > 1$, where encoding outperforms the break-even points.}}
\end{figure*}

As an illustration of the ET nature of the $\overline{X}$ gate, Fig. \ref{fig: ET X gate rotation illustrate} shows the spin populations of arsenic nucleus (spin $I=3/2$) over three complete rotations ($\phi \in [0,6\pi]$), initialized in the error-corrupted state $|\psi(0)\rangle = \sqrt{0.9}|0_L\rangle + \sqrt{0.1}|\mathbb{E}_0^1\rangle$, comprising a coherent superposition of the logical and its first-order error state, evolving under dephasing (see Methods for details).
The populations of $|\mathbb{E}_0^k\rangle$ and $|\mathbb{E}_1^k\rangle$ oscillate synchronously within their respective subspaces, while the logical subspace population $\mathcal{P}_0$ decreases over time, as expected as the dephasing drives transitions from $\mathcal{P}_0$ to $\mathcal{P}_1$.
Accumulation of dephasing errors further induces transitions from $\mathcal{P}_1$ into higher-order, uncorrectable subspaces. 
Nevertheless, the overall effect of noise on the logical information is suppressed relative to a non-ET gate implementation, confirming the advantage of the ET construction.

The main challenge in realizing the ET $\overline{X}(\phi)$ gate is the presence of third- and higher-order $\hat{I}_x$ terms in the gate Hamiltonian, which do not arise naturally from the donor spin interactions.
Hence, we discuss two possible strategies to realize this operation in practice.
In the first approach, the starting point is to use the GRF formalism and multi-tone driving to activate the SU(2) covariant rotation. 
Following the details in \cite{yu2025Schrodinger}, going to GRF and making appropriate choice for $\omega_k$, $\phi_k$, and setting $B_{d,k} = B_d$ in Eq. \eqref{eq: control hamiltonian} gives us the Hamiltonian $\hat{H}_{\text{GRF}} = \frac{\gamma_n B_1}{2}\hat{I}_x$. 
Then, we can go to a rotating frame defined by the operator $\hat{U}_{\text{rf}} = \exp\left[-it\left(b_{3}\hat{I}_x^3 + b_5\hat{I}_x^5 + \dots + b_{2I}\hat{I}_x^{2I}\right)\right]$, where $b_n$'s will be determined later. 
The resulting Hamiltonian is 
\begin{equation}
\hat{H}_{\text{GRF,rf}} = \frac{\gamma_n B_1}{2}\hat{I}_x + b^{(I)}_{3}\hat{I}_x^3 + b^{(I)}_5\hat{I}_x^5 + \dots + b^{(I)}_{2I}\hat{I}_x^{2I}\, .
\end{equation}
By comparing each of the above terms with Eq. \eqref{eq: ET X Hamiltonian linear decomposition}, one can easily determine what $b^{(I)}_n$'s are. 
More specifically, we first compare the first term, i.e., $\Omega c^{(I)}_1 = \gamma_{n}B_1/2$, to determine $\Omega$.  
Afterwards, it is straightforward to determine $b^{(I)}_n$'s from matching each subsequent term. 
The gate time for $\overline{X}(\frac{\pi}{2})$ is $T_{\pi/2} = \frac{\pi c^{(I)}_1}{\gamma_n B_1}$, which is just slightly slower than standard SU(2) rotation due to the factor $c^{(I)}_1$ being slightly larger than 1 and differing for different spin-$I$ systems. 
As a reference, for arsenic donor ($I=3/2$) and $B_d=0.1$ mT, the gate time $T_{\pi/2}$ is 0.684 ms for standard SU(2) rotation and 0.741 ms for the ET rotation.

The second approach relies on decomposing the unitary $\overline{X}(\pi/2)$ into a sequence of elementary operations which is possible due to the universality of coherent manipulation of the nuclear spins. 
Specifically, Theorem 1 in Ref.~\cite{merkel2009quantum} states that the set $\{\hat{I}_x, \hat{I}_y, h\}$ is sufficient to generate the Lie algebra $\mathfrak{su}(d_n)$ provided $h$ has non-zero overlap with at least one rank-2 irreducible spherical tensor.
There are many valid choices of $h$ that complete this generating set; one natural option is $h = \hat{I}_x^2$, corresponding to the one-axis twisting operation~\cite{ma2011quantum}, whose implementation in the context of donor nuclear spins has been discussed in Ref.~\cite{gupta2024robust}. 
More generally, the linear terms can be any two orthogonal spin operators (e.g. $\hat{I}_x, \hat{I}_z$), and the quadratic term is not restricted to the $x$-direction (but it cannot be $\hat{I}^2$). 
Taken together, this implies that control over both the quadrupole and magnetic terms, which has been demonstrated~\cite{fuentes2024navigating}, is necessary to implement $\overline{X}(\pi/2)$ via this approach.

Alternatively, the recently proposed {\it matrix-element modification (MEM)} protocol \cite{roy2025synthetic} offers another route to generating the nonlinear terms needed to complete the generating set.
Although Ref.~\cite{roy2025synthetic} develops the technique primarily in the context of superconducting circuits, it is easily adaptable for nuclear spin-based platforms, which we describe briefly. 
The adapted protocol requires an qubit ancilla (electron) and the application of a frequency comb on the qubit at frequencies $\omega_q + n\chi$ and phases $\phi_n$ for $n = 0, \dots,2I$, where $\omega_q$ is the qubit frequency and $\chi \propto A$ is the coupling strength between the nuclear spin and electron ancilla.
A two-tone drive is applied on the qudit (high-spin nucleus) system with frequencies $\omega_s + n\chi$ for $n=0,1$, where $\omega_s\propto \gamma_nB_0$ is the nuclear spin's frequency.
The resulting unitary is $\hat{U} = \exp(-i\theta \hat{M}_{\varphi})$ with the generator $\hat{M}_{\varphi} = e^{-i\varphi}\hat{M} + e^{i\varphi}\hat{M}^\dagger$, where $\hat{M} = \sum_{n=1}^{2I}\sqrt{n}\cos(\delta\phi_n/2)|n-1\rangle\langle n|$, and $\delta\phi_n = \phi_n - \phi_{n-1}$. 
The generator admits the decomposition $\hat{M} = \sum_{n=1}^{2I} c_n\left[\hat{I}_-,\hat{I}_z^n\right]$, where $n>1$ are the desired nonlinear terms. 
The combination of SU(2) rotations and a single nonlinear spin rotation is sufficient for universal SU($d_n$) control. 

The above approaches comes with their unique challenges.
The GRF approach requires $\mathcal{O}(d_n)$ cost associated with keeping track the phases of `virtual clocks', after which the rotating-frame Hamiltonian requires a classical diagonalization of $\hat{I}_x$ at $\mathcal{O}(d_n^3)$ cost per timestep via a similarity transformation. 
This may be inefficient to simulate classically. 
The second approach requires finding a unitary decomposition of $\overline{X}(\pi/2)$ and optimizing its sequence length and total operation time, for which no straightforward analytical method exists, to our knowledge. 
For the MEM protocol, two practical questions remain open: whether the adiabatic gate can be executed fast enough to outpace  dephasing, and whether the required coupling strengths are achievable within the dispersive coupling regime for the donor-ancilla system. 
Both warrant careful examination before the protocol can be considered viable for donor spin platforms.
We remark that the phase gates and amplitude-mixing gates are reminiscent of Clifford and non-Clifford (magic) gates respectively, where the latter typically demands a non-trivial resource overhead \cite{campbell2017road} --- a parallel that may inform future approaches to implementing $\overline{X}(\pi/2)$ efficiently. 


\subsection{ET gate performance with error correction}\label{sec: numerical performance}

Since the spin cat-$x$ codes satisfy the Knill-Laflamme conditions for rank-$r$ dephasing noise, an error correction protocol can be formulated following Ref.~\cite{knill1997theory}, which we discussed in full detail in SM Sec. C.
The protocol operates in three steps: 
(i) a decoding unitary $\hat{U}_\text{dec}$, comprising a logical Hadamard $\overline{H}$, an SU(2) rotation $e^{i\frac{\pi}{2}\hat{I}_y}$, and a SNAP gate, maps each error subspace $\mathcal{P}_k$ to a distinct orthogonal subspace spanned by $\{|\pm(I-k)\rangle\}$, making errors detectable by their spin projection; 
(ii) an electron ancilla initialized in $|\downarrow\rangle$ sequentially detects the $k$-th order error via conditional ESR $\pi$-pulses, with electron reset between steps; 
and (iii) conditional NMR $\pi$-pulses correct each detected error by mapping $|\pm(I-k)\rangle \to |\pm(I-k+1)\rangle$, proceeding from highest to lowest error order.
The protocol is measurement-free~\cite{pazsilva2010fault}, and upon completion the state is re-encoded via $\hat{U}_\text{en} =\overline{H} e^{i\frac{\pi}{2}\hat{I}_y}$ (i.e. in the reverse order to $\hat{U}_\text{dec}$ but without the SNAP gate) for continued operation.
Under ideal implementation (noiseless), this error correction protocol achieves the optimal recovery fidelity for spin cat codes \cite{gross2024hardware,chiesa2020molecular}.

To assess the performance of the proposed ET gates, we perform numerical simulations in which the encoded system evolves under the desired gate Hamiltonian, followed by a round of noisy error correction as outlined above. 
Using this protocol, we demonstrate the advantage of encoding in the spin cat-$x$ code over an unencoded qubit by evaluating the entanglement fidelity of the logical $\overline{X}$ and $\overline{CZ}$ gates. 
The numerical setup and simulation details are discussed in Methods and SM Sec. C.
Since the ET $\overline{Z}$ gate can be implemented virtually, we focus on the performance of the remaining two ET gates.

In Fig.~\ref{fig: ET X gate fidelities}, we show the entanglement infidelities for implementing $\overline{X}(\pi/2)$. 
Implementing the ET gate followed by error correction opens the possibility of beating the break-even line, which corresponds to the performance of an unencoded qubit.
Using the assumption that $B_d = 0.1$ mT, the gate time for $\overline{X}(\pi/2)$ is around 0.78 ms, which coupled with the reported $T_2^*$ for donor nuclei~\cite{morello2020donor,fuentes2024navigating}, we find that $\Gamma_n T_G$ falls around $10^{-2}$ to $10^{-1}$. 
This means that under the current experimental parameters, the gate time is near the optimal EC cycle time, where performing the encoding and correction have the most advantageous gain.
Notably, the ET gate performance is identical to the idle case where no gate is performed and the system dephases for the same duration, confirming that the ET construction introduces no additional logical error beyond free evolution.

While higher spins offer greater capacity of tolerating and correcting noise, it also incurs longer and more complex error correction circuit implementations. 
Consequently, higher-spin encodings yield only marginal gains at higher noise strengths, and their advantage diminishes at weak noise strengths or short correction cycles.
A similar trend is reported in Ref.~\cite{gross2024hardware}, although our results are distinguished by the use of realistic donor parameters.
This accounts for the comparatively weaker performance of the spin-7/2 donor, attributed to its slower NMR Rabi rates due to its smaller gyromagnetic ratio and hyperfine coupling compared to other donors (see Table~\ref{tab: group V donor parameters}). 

For additional comparison, we show the performance of the basic, non-ET logical $X$ gate
implemented with the Hamiltonian $\hat{H} \propto \hat{I}_x$, which remains
above the break-even line across all noise strengths and converges with the ET gate only in the small noise limit. 
This underscores the importance of implementing ET gates for achieving fault-tolerant performance below the break-even line.

Next, we study the performance of the ET $\overline{CZ}$ gate followed by noisy error correction for the four high-spin species, using an infidelity gain metric defined as the ratio of the gate infidelities of the unencoded qubit and the spin cat-$x$ encoding: $G = (1-\mathcal{F}_e^\text{qb})/(1-\mathcal{F}_e^\text{cat})$. 
$G > 1$ indicates that the spin cat-$x$ code outperforms the unencoded case.
We compare the cases where the ESR pulses are applied sequentially in $(I+1/2)^2$ steps, to a single application step with multi-tone driving.
As the $\overline{CZ}$ gate requires an electron ancilla, we plot the gain $G$ against electron dephasing rate $\Gamma_e$ and the dimensionless product of nuclear dephasing rate and gate time $\Gamma_n T_G$.
The results shown in Fig.~\ref{fig: ET $CZ$ gate performance complete} provide several physical insights.

Firstly, when electron dephasing dominates (top left of plots), both encoded and unencoded qubits perform identically (white regions) despite fast gate times and/or high nuclear coherence, showing that the coherence of the ancilla is the limiting factor in this regime. 
With smaller nuclear and electron dephasing rates (lower left of plots), the unencoded qubit performs better (blue region) and encoding becomes disadvantageous.
This is because  error correction takes finite time, during which dephasing can still occur.
Furthermore, the number of detection and correction steps scale with the rank $r$ of correctable errors  as $r = \lfloor I\rfloor$, which explains why this disadvantageous region generally increases in size as spin increases.

Secondly, intermediate values of $\Gamma_n T_G$ are where encoding becomes advantageous (red region), and gain is significantly better when ESR pulses are applied in a single, multi-tone drive step (top row) compared to sequentially (bottom row).
The performance of the simultaneous drive is also closer to the ideal performance.
This can be seen by taking a line cut at $\Gamma_e = 10^4$ s$^{-1}$ and comparing the results with those in Fig. \ref{fig: ET X gate fidelities}.
At this value of $\Gamma_e$, the maximum gain with sequential ESR pulses is approximately $G \approx 2$ for $I=3/2$, which requires the least number of pulse applications, and $G \approx 10^{1.48} \approx 30.20$ for $I=9/2$.

Finally, for the simultaneous pulse case, the size of the region where encoding is advantageous (red) increases with spin, while for the sequential pulse case, the size of this region decreases with spin. In both cases, there is a significant decrease in the size of these regions at $I=7/2$, which is in line with the findings from simulating $\overline{X}(\pi/2)$. 
This mirrors the previous simulation, where the bottleneck is the slower Rabi rates of the NMR pulses conditioned on the electron spin state during the error correction stages.

There are two key takeaways from our results. 
Firstly, there is considerable advantage in engineering multi-tone ESR capability for $\overline{CZ}$ gates. 
In this respect, recent work demonstrating two-tone ESR~\cite{collett2024configurable} is a promising development. 
Secondly, even when our code and error correction scheme does not correct for noise in the electron, we still find that break-even performance is observed and it primarily depends on the nuclear dephasing. 
With experimentally reported electron dephasing rates $\Gamma_e$ in the range of $10^{4}$ to $10^{5}~\mathrm{s}^{-1}$ \cite{asaad2020coherent}, current devices already operate in the regime where error correction offers an advantage, which improves further as electron coherence time increases.


\subsection{From ET gates to fault-tolerant components}\label{sec: FT preparation measurement EC}

The existence of a universal ET gate set is an important step towards fault-tolerance, as it enables the construction of fault-tolerant logical gates, measurement protocols and error correction procedures. 
Here, we adapt the conventional definition of fault-tolerance for multi-qubit codes to the spin code case (see SM Sec. D for formal definitions), and discuss the essential ingredients necessary to achieve full fault-tolerance for the spin cat-$x$ code.
High-fidelity logical state preparation, while not strictly fault-tolerant, can nonetheless be achieved via optimal control with high-fidelity, as we demonstrate below.

\subsubsection{Near fault-tolerant state preparation}

Following Refs.~\cite{asaad2020coherent,fuentes2024navigating}, the nuclear spin can be initialized in any target eigenstate $|m_I^{\text{target}}\rangle$ via two approaches: an adiabatic frequency sweep through the ESR frequencies combined with spin-selective tunneling readout and NMR $\pi$-
pulses, or spin pumping via flip-flop (FF) transitions, which requires no calibrated nuclear pulses, measurements, or feedback.

Taking the ground $|I,I\rangle$ state as the natural starting point, the task is then to prepare an arbitrary logical state $|\psi_L\rangle = \alpha|0_L\rangle + \beta|1_L\rangle$ within the codespace.
This can be done by performing a $\pi/2$ covariant rotation around the $y$-axis, which brings $|I,I\rangle$ to $x$-basis spin coherent state $|I,I\rangle_x$. Being the dual of the logical codewords, $|I,I\rangle_x$ already affords protection against dephasing errors, and any target logical state $|\psi_L\rangle$ is then reachable via the ET gates constructed above. 
This has been demonstrated recently in Ref.~\cite{yu2025Schrodinger} with fidelity of approximately $93\%$. 

An important constraint worth noting is that because the initial state lies outside the code as well as the correctable error space, the initialization into logical code words cannot be made error-transparent.
However, the fidelity can be improved by optimizing the covariant rotation pulse using optimal control algorithms such as GRAPE \cite{khaneja2005optimal} or Krotov \cite{krotov1993global}. 
Using \texttt{qocttools} \cite{castro2022optimal,castro2024qoct}, we simulated the preparation of $|I,I\rangle_x$ and obtained fidelities $\gtrsim 99.9\%$ across all donor spins under dephasing noise.
We also refer the reader to Refs.~\cite{peham2026optimizing,chen2026faulttolerant} for recent developments in optimizing logical state preparation in cat codes.

The simulated preparation fidelity of $\gtrsim 99.9\%$ can be improved  further via optimal control to meet even the most stringent fault-tolerance thresholds, though the preparation step does not satisfy the error-transparency conditions, and hence falls short of full fault-tolerance in the strict sense.

\subsubsection{Error-transparent measurement}\label{subsec: ET measurement}

Unlike state preparation, logical $Z$-basis measurement can be made fully error-transparent, owing to the disjoint support between the $|\mathbb{E}_0^k\rangle$ and $|\mathbb{E}_1^k\rangle$ subspaces.
To this end, we construct the ET $Z$-basis measurement by 
\begin{equation}\label{eq: ET Z measurement operators}
\mathcal{M}_{\overline{Z}}^{\texttt{ET}} = (+1)\hat{M}^{Z,\texttt{ET}}_0 + (-1)\hat{M}^{Z,\texttt{ET}}_1\, ,\quad \hat{M}^{Z,\texttt{ET}}_\mu = \sum_{k=0}^r |\mathbb{E}_\mu^k\rangle \langle \mathbb{E}_\mu^k | \, ,
\end{equation}
where $\hat{M}^{Z,\texttt{ET}}_\mu$ is the ET measurement operator corresponding to the outcome $\mu$. 
These measurement operators are error-transparent in the sense that any state in $\text{span}\{|\mathbb{E}_\mu^0\rangle,|\mathbb{E}_\mu^1\rangle,\dots,|\mathbb{E}_\mu^r\rangle\}$ is mapped to the correct outcome $\mu$ regardless of which error sector it occupies.
This is in contrast to the basic logical $Z$-measurement with operators $\hat{M}_{\mu}^{Z,\texttt{DUAL}} = |\mu_L\rangle \langle \mu_L|$, which projects onto the logical codespace and thus fails in the presence of uncorrected errors. 
We show in SM Sec. D.1 that the ET measurement here is fully FT.

Although the above operators appear complicated to realize, they reduce to the spin parity operator 
\begin{eqnarray}
\hat{\Pi} &=& \sum_{m_I = -I}^I (-1)^{I+m_I}|I,m_I\rangle \langle I,m_I| \nonumber \\
&=& (+1)\underbrace{\left(\sum_{m_I\in \text{even}}|I, m_I\rangle \langle I, m_I |\right)}_{\hat{M}^{Z,\texttt{ET}}_0} \nonumber \\ 
& & \qquad + (-1)\underbrace{\left(\sum_{m_I\in \text{odd}}|I, m_I\rangle \langle I, m_I |\right)}_{\hat{M}^{Z,\texttt{ET}}_1}\, ,
\end{eqnarray}
where the even states $\{|I,-I\rangle, |I,-I+2\rangle, \dots |I, I-1\rangle\}$ carry $+1$ parity, and the odd states $\{|I,I\rangle , |I,I-2\rangle, \dots ,|I,-I+1\rangle \}$ carry $-1$ parity.
This structure is made transparent by the Holstein-Primakoff mapping~\cite{holstein1940field} following $|I,m_I\rangle \rightarrow |n\rangle$ with $n = I + m_I$, under which even/odd parity corresponds directly to even/odd Fock state occupation. The $0/1$ outcome of the ET logical $Z$-measurement thus reads out the even/odd parity of the nuclear spin population, which has been demonstrated experimentally in Ref.~\cite{yu2025Schrodinger}.
Measurement in the dual basis, i.e., the logical $X$-basis, is obtained by conjugating the ET $Z$-measurement with an ET Hadamard gate, decomposed as $\overline{H} = \overline{Z}(\frac{\pi}{2})\overline{X}(\frac{\pi}{2})\overline{Z}(\frac{\pi}{2})$ from the ET gate set above. 

\subsubsection{Knill's error correction by teleportation}\label{subsec: knill EC}

The error correction protocol discussed in the previous section (also in SM Sec. C) is optimal in its capacity to correct dephasing errors up to order $r$ when it is ideal.
However, it is not strictly fault-tolerant, as it involves operations that are not error-transparent --- specifically SU(2) rotations and population transfer via ESR and NMR pulses. 
A further complication arises when noise on the electron ancilla is taken into account, since this is not addressed by the protocol. 
We therefore seek a fault-tolerant alternative, focusing on noise in the donor spins.

A natural framework is Knill's error correction (Knill-EC)~\cite{gottesman2009introduction}, also known as \textit{telecorrection} owing to its teleportation-based circuit structure. 
Knill-EC is well-suited to our setting because its modular structure allows different encodings, gates, and measurements to be optimized independently across the circuit. 
The bosonic adaptation for cat codes serves as our starting point, and we construct a spin cat-$x$ code variant in which every component of the circuit is realized by an error-transparent operation.

\begin{figure}
\centering
\includegraphics[width=1\linewidth]{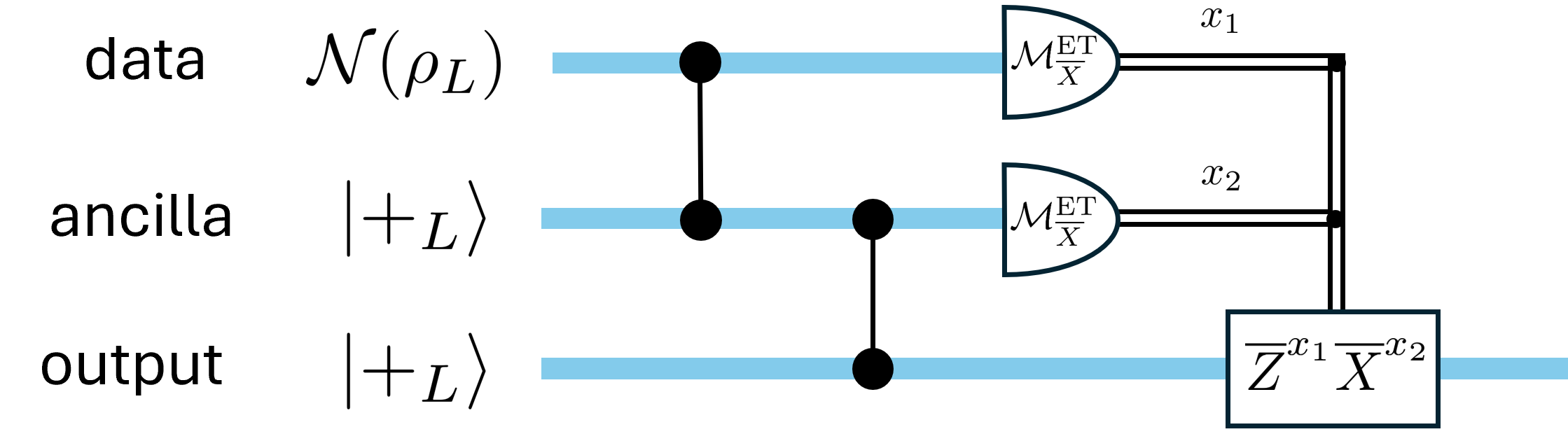}
\caption{\justifying{Knill error correction (Knill-EC) circuit based on teleportation. All the rails of the circuit are assumed to be the same high-spin system (denoted with blue thick lines) that is encoded in spin cat-$x$ code. The data rail contains a noisy logical state represented by the error channel $\mathcal{N}$ acting on $\rho_L$, while the ancilla and output rail are initialized in the logical dual state $|+_L\rangle$. The remaining circuit components consists of ET version of $\overline{CZ}$ gates, logical dual basis measurements, and appropriate Pauli gates depending on the measurement outcomes.}}
\label{fig: telecorrection circuit}
\end{figure}

The resulting circuit is shown in Fig.~\ref{fig: telecorrection circuit}, and consists of three rails --- data, ancilla, and output --- all realized as high-spin systems encoded in the spin cat-$x$ code. 
The data rail carries the corrupted logical state, while the ancilla and output rails are initialised in $|+_L\rangle = \frac{1}{\sqrt{2}}(|0_L\rangle + |1_L\rangle)$. 
Even though the ancilla rail could have a different dimension from the data rail, having all rails of the same type is advantageous, as the required $\overline{CZ}$ gates between them can be implemented in ET form, preventing detrimental error propagation.
ET logical $X$-basis measurements on the data and ancilla rails then serve a dual purpose: they distinguish between logical codewords and error words, and teleport the logical information to the output rail.
The classical outcomes $(x_1, x_2) \in \{0,1\}^2$ determine the recovery operation $\overline{Z}^{x_1}\overline{X}^{x_2}$ applied to the output, where $\overline{Z} = \overline{Z}(\pi)$ and $\overline{X} = \overline{X}(\pi)$ are ET logical Pauli operators. 
In this construction, the Knill-EC circuit is composed entirely of error-transparent operations.

\begin{figure*}
\centering
\begin{subfigure}{0.48\textwidth}
\centering
  \includegraphics[width=0.99\linewidth]{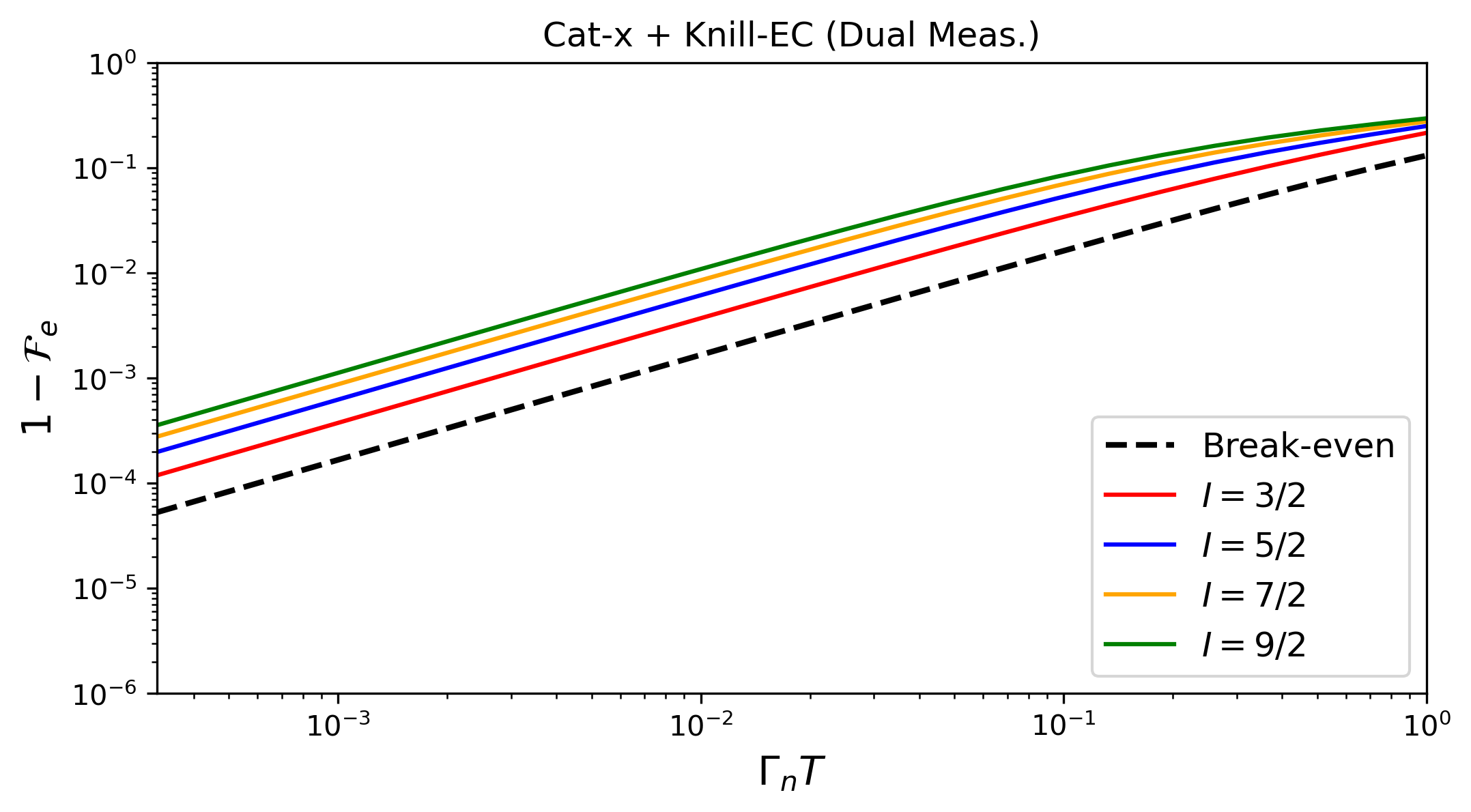}
  \caption{}
  \label{fig: Knill-EC fidelities (dual-dual)}
\end{subfigure}
\begin{subfigure}{0.48\textwidth}
\centering
  \includegraphics[width=0.99\linewidth]{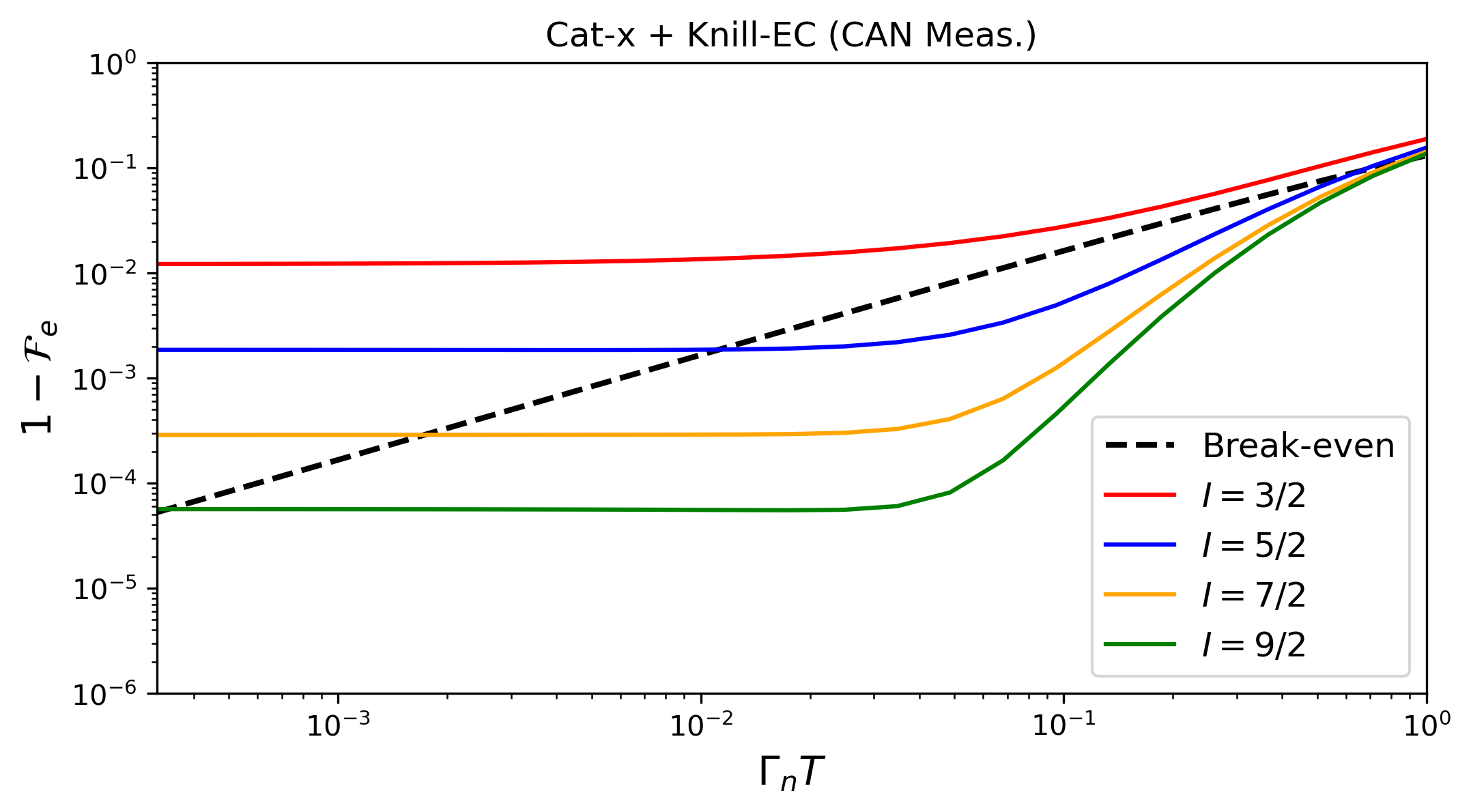}
  \caption{}
  \label{fig: Knill-EC fidelities (CAN-dual)}
\end{subfigure}
\begin{subfigure}{0.48\textwidth}
\centering
  \includegraphics[width=0.99\linewidth]{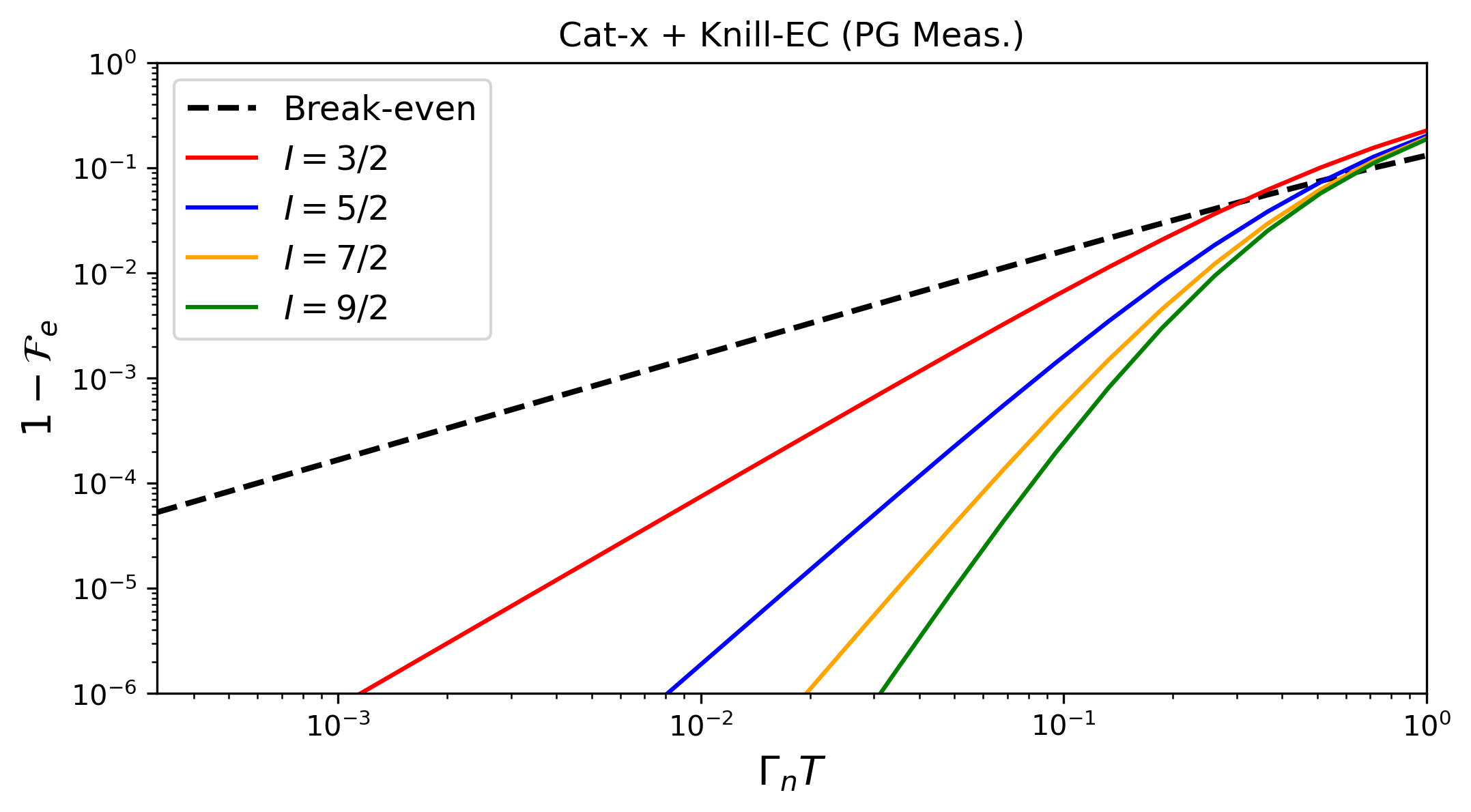}
  \caption{}
  \label{fig: Knill-EC fidelities (PG-dual)}
\end{subfigure}
\begin{subfigure}{0.48\textwidth}
\centering
  \includegraphics[width=0.99\linewidth]{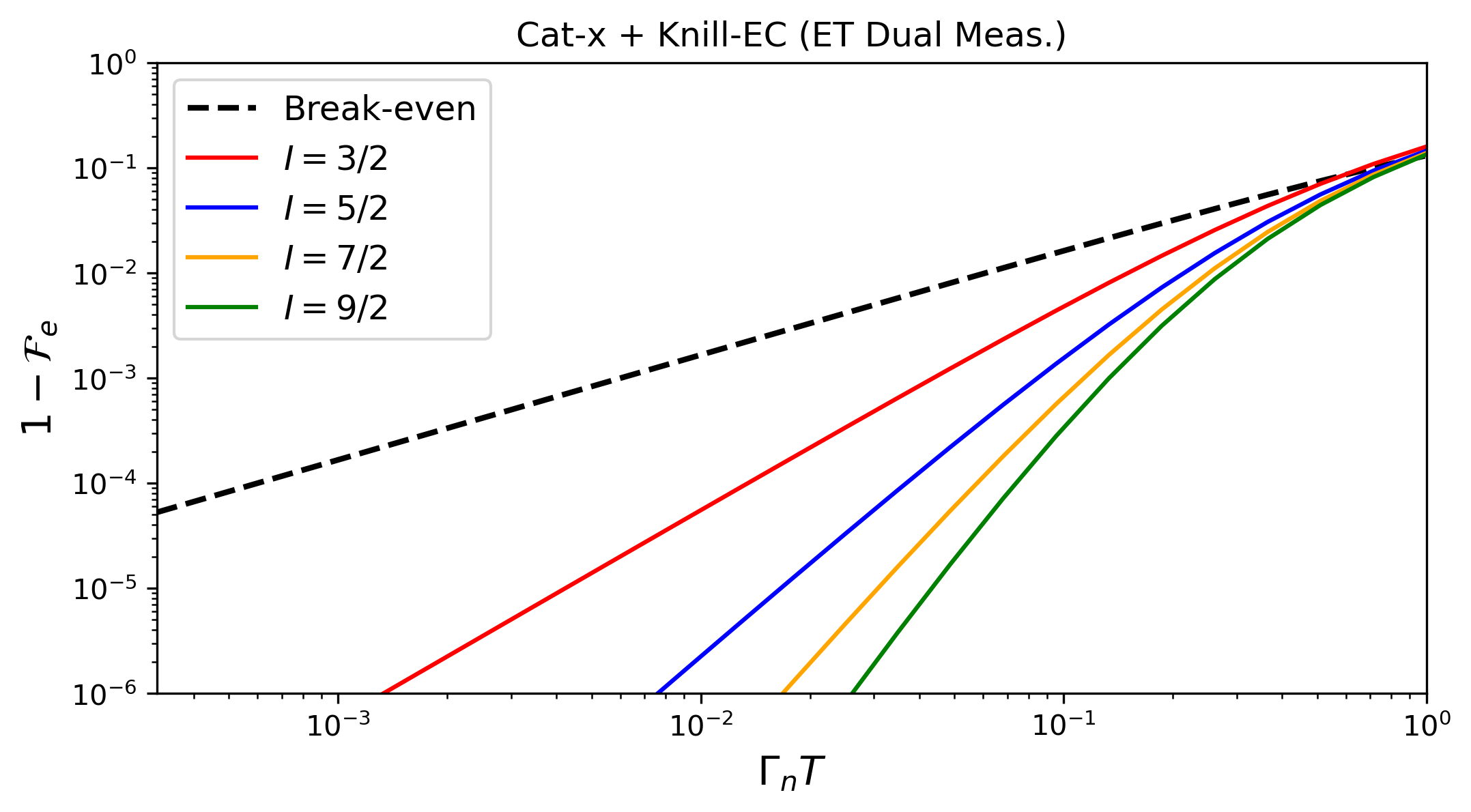}
  \caption{}
  \label{fig: Knill-EC fidelities (ET-dual)}
\end{subfigure}
\caption{\justifying{Knill-EC performance of several spin systems with various dual basis measurement setups in the data rail: (a) basic logical measurement (\texttt{DUAL}), (b) canonical phase measurement (\texttt{CAN}), (c) pretty good measurement (\texttt{PG}), and (d) error-transparent (\texttt{ET}). The simulation is performed with noise only in the input; error-free implementations of Knill-EC are assumed. Break-even lines correspond to the performance of the uncorrected qubit acting as the benchmark. The numerical performance of (c) and (d) is very similar to the performance of the ideal EC when the recovery operation is noise-free.}}
\label{fig: Knill-EC fidelities (various)}
\end{figure*}

The fault-tolerance of this circuit can be analysed at the circuit level following Ref.~\cite{my2025circuit}. 
Examining fault locations at every step, one finds that the ET properties of the gates and measurements prevent correctable errors from propagating into uncorrectable ones. 
It can thereby be shown that the circuit satisfies an adapted version of the fault-tolerance criteria of Ref.~\cite{aliferis2006quantum} for error correction gadgets, under the assumption of nuclear dephasing noise in the rails. 
The formal definition and proof are given in SM Sec. D.2.

In practice, however, the ET $\overline{CZ}$ gate requires an electron ancilla to mediate the inter-donor coupling, so at least two electrons are needed in the full Knill-EC circuit.
The circuit therefore does not achieve full fault-tolerance when electron noise is also present -- a limitation we return to in the Discussion.

To assess the error correction capacity of the Knill-EC gadget, we simulate its performance ideally following the setup of Refs.~\cite{grimsmo2020quantum, hillmann2022performance}: noise acts only on the input state, while the ancilla rails are error-free and all gates and measurements are ideal. We compare four realizations of the logical $X$-basis measurement in the data rail -- basic dual (\texttt{DUAL}), canonical phase (\texttt{CAN}) \cite{grimsmo2020quantum}, pretty good (\texttt{PG}) \cite{hausladen1994prettygood}, and error-transparent (\texttt{ET}) -- with a basic dual measurement fixed in the ancilla rail throughout, which is optimal given the error-free ancilla assumption. Details of the measurement constructions and simulation parameters are given in SM Sec. E.

The results are shown in Fig.~\ref{fig: Knill-EC fidelities (various)}. The \texttt{DUAL} measurement fails to surpass the break-even line at any noise strength. The \texttt{CAN} measurement crosses break-even for sufficiently large spin but plateaus at low noise.
This is due to the nature of the \texttt{CAN} measurement which cannot deterministically distinguish logical states, thus imposing a bound on the maximum fidelity.
In contrast, both \texttt{PG} and \texttt{ET} measurements achieve performance nearly identical to the ideal EC for spin codes~\cite{chiesa2020molecular, gross2024hardware}. While \texttt{PG} is known to be near-optimal, it requires prior knowledge of the noise profile, which may not be available in practice. The \texttt{ET} measurement requires no such knowledge and is implementable within our ET gate set. That it matches ideal EC performance underscores the strength of error-transparent operations and their suitability for Knill-EC and potentially other fault-tolerant schemes.


\section{Discussion}\label{sec: discussion}

Recent progress in the coherent manipulation and long coherence times of high-spin donors in silicon motivates the exploration of novel quantum error correcting codes in this platform. Toward the goal of fault-tolerant quantum computing, we have proposed a universal set of error-transparent gates for spin cat-$x$ codes of spin $I$, protecting against the dominant nuclear dephasing noise up to rank $r = \lfloor I \rfloor$. Error-transparent gates allow errors to occur at any point during gate execution without erroneously transforming the error words, so that subsequent error correction recovers more of the computational data and yields better logical fidelity. Through numerical simulation, we find that ET gates offer significant protection against dephasing compared to basic logical gate implementations. Even with electron ancilla noise included, sweet spots exist at which break-even is achieved, and the optimal error correction cycle can be identified.

Building on the ET gate set, we have further discussed progress toward full fault-tolerance, covering state preparation, measurement, and error correction. In particular, we show that spin parity measurement is error-transparent, and that the Knill-EC protocol constructed entirely with ET operations matches the performance of the ideal error correction circuit.

Several limitations warrant further investigation. 
First, several ET operations require an electron ancilla, which is itself susceptible to dephasing. When electron dephasing is significant, logical fidelities degrade and the advantage of error correction diminishes. One avenue for mitigation is to employ a higher-dimensional ancilla together with path-independent or ET control schemes, as discussed in Refs.~\cite{xu2024fault, wetherbee2025mathematical}.
Second, while relaxation errors are negligible in the current setting, it is worth developing error correction and fault-tolerance formalisms against more general noise. 
A natural route is to concatenate the spin cat code with a repetition code: Ref.~\cite{omanakuttan2024fault} showed that three copies of a spin cat code can correct general errors up to rank $K = \lfloor\frac{2I-1}{2}\rfloor$, and this framework would also allow a threshold analysis for the concatenated code.
Third, our Knill-EC analysis has so far assumed a noiseless gadget. 
The next step is a realistic circuit-level simulation in which noise can occur at any operation, following the extended rectangle configuration of Refs.~\cite{aliferis2006quantum, my2025circuit}. This is numerically demanding for higher spin systems with $I \geq \frac{7}{2}$, as encountered in our ET $\overline{CZ}$ gate simulations. Finally, a hybrid error correction scheme in the spirit of Steane~\cite{steane1997active}, constructed from ET components, is a promising direction: the bosonic analogue studied in Refs.~\cite{grimsmo2020quantum, my2025circuit} was shown to outperform Knill-EC when optimized. 

The implementation of the logical ET $\overline{X}(\pi/2)$
gate remains the most technically demanding component of the universal ET gate set, with each proposed approach carrying its own open challenges, as discussed earlier. 
Resolving these challenges represents a key near-term milestone toward a fully hardware-efficient implementation of fault-tolerant spin cat-$x$ code operations.
We leave these directions as future work.
The open questions identified here are rich and non-trivial, but their resolution would chart a concrete path toward full fault-tolerant quantum computing in the donor-in-silicon architecture.


\section{Methods}

\subsection{Hamiltonian of ET $\overline{X}$ gate}\label{method: ET X gate}

In the main text, we discuss the Hamiltonian form for the ET $\overline{X}$ gate for spin codes. 
Here, we show the derivations more explicitly:
\begin{small}
\begin{align}
\hat{H}_{\overline{X}}/\Omega & = \frac{1}{2}\sum_{k=0}^{r} \left(|\mathbb{E}_{1}^k\rangle \langle \mathbb{E}_{0}^k| + |\mathbb{E}_{0}^k\rangle \langle \mathbb{E}_{1}^k|\right) \\
& = \frac{1}{2}\Bigg[\underbrace{\left(|\mathbb{E}_{1}^0\rangle \langle \mathbb{E}_{0}^0| + |\mathbb{E}_{0}^0\rangle \langle \mathbb{E}_{1}^0|\right)}_{\text{codespace}} + \underbrace{\left(|\mathbb{E}_{1}^1\rangle \langle \mathbb{E}_{0}^1| + |\mathbb{E}_{0}^1\rangle \langle \mathbb{E}_{1}^1|\right)}_{\text{1st error subspace}} + \dots  \nonumber \\
& \quad + \underbrace{\left(|\mathbb{E}_{1}^r\rangle \langle \mathbb{E}_{0}^r| + |\mathbb{E}_{0}^r\rangle \langle \mathbb{E}_{1}^r|\right)}_{r\text{-th error subspace}}\Bigg] \\
& = \frac{1}{2}\Big[|I,I\rangle_x\langle I,I| - |I,-I\rangle_x\langle I,-I|\Big] \nonumber \\
& \quad  + \frac{1}{2}\Big[|I,I-1\rangle_x\langle I,I-1| - |I,-I+1\rangle_x\langle I,-I+1|\Big] \nonumber \\
& \quad + \dots + \frac{1}{2}\Big[|I,\frac{1}{2}\rangle_x\langle I,\frac{1}{2}| - |I,-\frac{1}{2}\rangle_x\langle I,-\frac{1}{2}|\Big] \\
&= \frac{1}{2}\sum_{k=0}^r |I,I-k\rangle_x\langle I,I-k| - |I,-I+k\rangle_x\langle I,-I+k| 
\end{align}
\end{small}
As discussed in the main text, we can equate the above with linear combinations of $\hat{I}_x^n$ following Eq. \eqref{eq: ET X Hamiltonian linear decomposition}. 
To determine the coefficients $c_n^{(I)}$ for spin $I$ system, we are solving an $r+1 = \lfloor I \rfloor + 1$ linear equations:
\begin{small}
\begin{equation}
\begin{cases}
Ic_1^{(I)} + I^3 c_3^{(I)} + \dots + I^{2I}c^{(I)}_{2I} & = \frac{1}{2} \\
(I-1)c_1^{(I)} + (I-1)^3 c_3^{(I)} + \dots + (I-1)^{2I}c^{(I)}_{2I} & = \frac{1}{2} \\
(I-2)c_1^{(I)} + (I-2)^3 c_3^{(I)} + \dots + (I-2)^{2I}c^{(I)}_{2I} & = \frac{1}{2} \\
& \vdots \\
\frac{1}{2}c_1^{(I)} + \left(\frac{1}{2}\right)^3 c_3^{(I)} + \dots + \left(\frac{1}{2}\right)^{2I}c^{(I)}_{2I} & = \frac{1}{2}
\end{cases}
\end{equation}
\end{small}
The above linear equations can be solved easily either analytically or numerically. 
We show the explicit decompositions for spin $\frac{3}{2}\leq I \leq \frac{9}{2}$: 
\begin{widetext}
\begin{align}\label{eq: spin cat-$x$ ET X hamiltonian} \hat{H}_{\overline{X}}/\Omega = \begin{cases}
\frac{13}{12}\hat{I}_x - \frac{1}{3}\hat{I}_x^3\, ,\quad & I=\frac{3}{2} \\
\frac{1067}{960}\hat{I}_x - \frac{11}{24}\hat{I}_x^3 + \frac{1}{20}\hat{I}_x^5\, ,\quad & I=\frac{5}{2} \\
\frac{30251}{26880}\hat{I}_x - \frac{301}{576}\hat{I}_x^3 + \frac{61}{720}\hat{I}_x^5 - \frac{1}{252}\hat{I}_x^7\, ,\quad & I=\frac{7}{2} \\
\frac{5851067}{5160960}\hat{I}_x - \frac{46573}{82944}\hat{I}_x^3 + \frac{7501}{69120}\hat{I}_x^5 - \frac{97}{12096}\hat{I}_x^7 + \frac{1}{5184} \hat{I}_x^9\, ,\quad & I=\frac{9}{2}
\end{cases}
\end{align}
\end{widetext}


\subsection{Numerical simulation}

In this section, we provide details for evaluating the numerical performance of ET gates and error correction protocol. 
The effect of the error channel is obtained by solving the master equation described by
\begin{equation}\label{eq: master equation}
\frac{d\rho}{dt} = -i\left[\hat{H}(t), \rho\right] +  \Gamma_n\mathcal{D}[\hat{I}_z]\rho + \frac{\Gamma_e}{2}\mathcal{D}[\hat{\sigma}_z]\rho\, ,
\end{equation}
where $\hat{H}(t)$ is the gate Hamiltonian, $\Gamma_n$ ($\Gamma_e$) is the nucleus (electron) dephasing rate, and $\mathcal{D}[A](\bullet) = A\bullet A^\dagger - \frac{1}{2}\{A^\dagger A, \bullet\}$ is the Linbladian dissipative channel. 
Since we only consider dephasing noise, the error operators are $\hat{I}_z$ and $\hat{\sigma}_z$ for the nucleus and electron system, respectively. 
The state $\rho$ is the density matrix corresponding to the nucleus-electron system that can be generalized easily depending on the system configuration.
As an example, for the CZ implementation, $\rho$ corresponds to $\mathrm{n}_1-\mathrm{n}_2-e$ (2 donors + electron) system, and the error model is extended for each system independently.
In this case, we have $\Gamma_{n_1}\mathcal{D}[\hat{I}_z^{(1)}]\rho + \Gamma_{n_2}\mathcal{D}[\hat{I}_z^{(2)}]\rho$. 
We numerically solve the simulation for time $T_G$, the time needed to achieve the unitary transformation with respect to the implemented gate. 

The performance metric is calculated using the process (entanglement) fidelity \cite{hashim2025practical} given by
\begin{equation}\label{eq: entanglement fidelity}
\mathcal{F}_e = \sum_{M\in \{P,X,Y,Z\}} \frac{1}{8}\tr{M_L \mathcal{R}\circ \mathcal{N} \circ \mathcal{U}^\dagger(M_L)}\, ,
\end{equation}
where $M_L\in \{P_L,X_L,Y_L,Z_L\}$ are the logical projection and Pauli operators.
The sequence of channels in Eq. \eqref{eq: entanglement fidelity} is given by a perfect reverse unitary evolution $\mathcal{U}^\dagger(\bullet) = U^\dagger \bullet U$, $U = \exp(-iT_G \hat{H})$, followed by noisy time-evolution given by the channel $\mathcal{N}$ as a result of solving Eq. \eqref{eq: master equation}, and lastly recovery operation $\mathcal{R}$.
The recovery operation $\mathcal{R}$ is simulated in finite-time using a piecewise-constant control Hamiltonian following the sequence of operations laid out in Fig. \ref{fig: ideal EC protocol general}.
Moreover, we run it in parallel with the presence of nuclear dephasing rate (varies) and electron dephasing rate of $\Gamma_e = 10^4$ s$^{-1}$ in the detection and correction steps. 
The simulation is conducted as realistic as possible by considering physical parameters of the donor nucleus system (see Results and Table \ref{tab: group V donor parameters}) with applied magnetic field strength of $B_d = 0.1$ mT.
This puts the driving in the slow-driving regime and enables implementation in the rotating wave approximation. 
As a benchmark, we compare the performance against $\ce{^{31}P}$ donor (spin $I=1/2$) that acts as the uncorrected qubit system, i.e., as the break-even line. 


\section*{Acknowledgment}

This research is supported by the Ministry of Education, Singapore, under its Academic Research Fund Programme (MOE-T2EP50222-0017,  RG154/24 and RG182/25), NTU SPMS Collaborative Research Award, and National Research Foundation, Singapore and A*STAR under its Quantum Engineering Programme 2.0 (NRF2021-QEP2-02-P07). 
K. E. J. G. acknowledges the funding from the National Research Foundation, Singapore through the National Quantum Office, hosted in A*STAR, under its Centre for Quantum Technologies Funding Initiative (S24Q2d0009).
We would like to acknowledge the High Performance Computing Centre of Nanyang Technological University Singapore, for providing the computing resources, facilities, and services that have contributed to this work.

\section*{Data Availability}

The datasets generated during and/or analyzed in the current study are available from the corresponding authors on reasonable request.

\section*{Code Availability}

The codes used in performing the numerical simulations are available from the corresponding authors on reasonable request. 

\section*{Competing Interests}

The authors declare no competing interests.

\normalem

\bibliography{ref}

\ULforem

\newpage

\onecolumngrid

\begin{appendix}

\begin{center}
{\large \textbf{Supplementary Material: Towards fault-tolerance with universal phase-error-transparent gates for high-spin cat codes}}
\end{center}

\section{A. System parameters}\label{appendix: system details}

We show in Table \ref{tab: group V donor parameters} the physical parameters that are used in the numerical simulation.
These parameters are taken directly from \cite{morello2020donor}. 

\begin{table}[!htb]
\setlength{\tabcolsep}{9pt} 
\renewcommand{\arraystretch}{1.2}  
\centering
\caption{\justifying{Hamiltonian parameters for group-V donors in silicon. Electron gyromagnetic ratio, $\gamma_e/2\pi \approx 27.97$ GHz/T. We have assumed a simple model for the quadrupole interaction strength such that they are non-zero only in the $Q_{zz}$ component and take values that produces static quadrupole splitting, $f_Q \approx Q_{xx} + Q_{yy} - 2Q_{zz}$, that is in the same order as experimental findings. Namely, $f_Q = 50$ kHz for \ce{^{31}As} \cite{gross2024hardware}, and $f_Q = -66$ kHz for \ce{^{123}Sb}~\cite{asaad2020coherent}. To our knowledge, no experimental works have reported the $f_Q$ values for \ce{^{121}Sb} and \ce{^{209}Bi}, so we assumed they take the same value as \ce{^{123}Sb}.}
}
    \begin{tabular}{ccccccc}
    \hline \hline
    \multicolumn{1}{c}{Donor} & \multicolumn{1}{c}{$I$} & \multicolumn{1}{c}{$\gamma_n/2\pi$} & \multicolumn{1}{c}{$A/2\pi$} & \multicolumn{1}{c}{$q_n/2\pi$}  & \multicolumn{1}{c}{$Q_{zz}/2\pi$} \\ 
          &       & \multicolumn{1}{c}{[MHz/T]} & \multicolumn{1}{c}{[MHz/T]} & \multicolumn{1}{c}{[$10^{-28}$ m$^2$]} & \multicolumn{1}{c}{[kHz]} \\
    \hline
    \ce{^{31}P}   & 1/2   & 17.26 & 117.53 & - & - \\
    \ce{^{75}As}  & 3/2   & 7.31  & 198.35 & 0.314 & -25\\
    \ce{^{121}Sb} & 5/2   & 10.26 & 186.8 & [-0.36,-0.54] & 33 \\
    \ce{^{123}Sb} & 7/2   & 5.55  & 101.52 & [-0.49,-0.69] & 33 \\
    \ce{^{209}Bi} & 9/2   & 6.96  & 1475.4 & [-0.37,-0.77] & 33 \\
    \hline \hline
    \end{tabular}%
  \label{tab: group V donor parameters}%
\end{table}%

\section{B. Additional details on spin cat codes}\label{appendix: code and EC scheme details}

The spin cat code can be defined more generally as an equal superposition of spin coherent states \cite{radcliffe1971properties,arecchi1972atomic} aligned along the $\hat{\mathbf{v}}$-axis, $\hat{\mathbf{v}} \in S^2$, with codewords
\begin{eqnarray}
|0_{\text{cat}}\rangle_{\hat{\mathbf{v}}} &=& \frac{1}{\sqrt{2}}\Big(|I,+I\rangle_{\hat{\mathbf{v}}} + |I,-I\rangle_{\hat{\mathbf{v}}} \Big)\, , \nonumber \\
|1_{\text{cat}}\rangle_{\hat{\mathbf{v}}} &=& \frac{1}{\sqrt{2}}\Big(|I,+I\rangle_{\hat{\mathbf{v}}} - |I,-I\rangle_{\hat{\mathbf{v}}} \Big)\, ,
\end{eqnarray}
where $(\hat{\mathbf{v}}\cdot \hat{\mathbf{I}})|I,m\rangle_{\hat{\mathbf{v}}} = m|I,m\rangle_{\hat{\mathbf{v}}}$, $m\in\{I,I-1,\dots,-I\}$.
Let $\hat{\mathbf{v}} = [\cos\theta\sin\phi,\sin\theta\sin\phi,\cos\phi]$, then there exists a rotation such that $|I,I\rangle_{\hat{\mathbf{v}}} = \exp\left[-i\Theta(\hat{I}_x\cos\Phi - \hat{I}_y\sin\Phi)\right]|I,I\rangle$. 
Note that $|I,-I\rangle_{\hat{\mathbf{v}}} = |I,I\rangle_{-\hat{\mathbf{v}}}$.
We choose $\hat{\mathbf{v}} = \hat{\mathbf{x}} = [1,0,0]$ as the basis for our spin cat encoding, referred to as spin cat-$x$ code. 
The explicit codewords and errorwords for the spin cat-$x$ code are shown in Table \ref{tab: spin cat-$x$ error basis}. 
It is clear that $\text{supp}_{m_I}(|\mathbb{E}_{\mu}^k\rangle) = \text{supp}_{m_I}(|\mu_L\rangle)$ $\forall\, \mu,k$, where we have use the definition $\text{supp}_{m_I}(|\psi\rangle) = \{|I,m_I\rangle\,\vert\,\langle I, m_I|\psi\rangle\neq 0\}$.

\begin{table*}[!htb]
\renewcommand{\arraystretch}{1.5}  
\centering
  \caption{Errorwords $\{\mathbb{E}_\mu^k\rangle\}$ for spin cat-$x$ code in $z$ basis representation for $\frac{3}{2}\leq I \leq \frac{9}{2}$.}
    \begin{tabular}{|c|c||cccccccccc|}
    \hline 
     $I$ & Code  & $|+\frac{9}{2}\rangle$  & $|+\frac{7}{2}\rangle$  & $|+\frac{5}{2}\rangle$  & $|+\frac{3}{2}\rangle$  & $|+\frac{1}{2}\rangle$  & $|-\frac{1}{2}\rangle$  & $|-\frac{3}{2}\rangle$  & $|-\frac{5}{2}\rangle$  & $|-\frac{7}{2}\rangle$  & $|-\frac{9}{2}\rangle$\\
    \hline
    \multirow{4}[4]{*}{$\frac{3}{2}$} & $|\mathbb{E}_{0}^{0}\rangle$ & & & & 0 & $\frac{\sqrt{3}}{2}$ & 0 & $\frac{1}{2}$ & & &  \\
    & $|\mathbb{E}_{0}^{1}\rangle$ & & & & 0 & $\frac{1}{2}$ & 0 & $-\frac{\sqrt{3}}{2}$ & & &  \\
    & $|\mathbb{E}_{1}^{0}\rangle$ & & & & $\frac{1}{2}$ & 0 & $\frac{\sqrt{3}}{2}$ & 0     & & &  \\
    & $|\mathbb{E}_{1}^{1}\rangle$ & & & & $\frac{\sqrt{3}}{2}$ & 0 & $-\frac{1}{2}$ & 0     & & &  \\
    \hline
    \multirow{6}[6]{*}{$\frac{5}{2}$} & $|\mathbb{E}_{0}^{0}\rangle$ & & & 0 & $\frac{\sqrt{5}}{4}$ & 0 & $\frac{\sqrt{10}}{4}$ & 0 & $\frac{1}{4}$ & & \\
    & $|\mathbb{E}_{0}^{1}\rangle$ & & & 0 & $\frac{\sqrt{9}}{4}$ & 0 & $-\frac{\sqrt{2}}{4}$ & 0 & $-\frac{\sqrt{5}}{4}$ & & \\
    & $|\mathbb{E}_{0}^{2}\rangle$ & & & 0 & $\frac{\sqrt{2}}{4}$ & 0 & $-\frac{\sqrt{4}}{4}$ & 0 & $\frac{\sqrt{10}}{4}$ & & \\
     & $|\mathbb{E}_{1}^{0}\rangle$ &  & & $\frac{1}{4}$ & 0 & $\frac{\sqrt{10}}{4}$ & 0 & $\frac{\sqrt{5}}{4}$ & 0 & & \\
     & $|\mathbb{E}_{1}^{1}\rangle$ &  & & $\frac{\sqrt{5}}{4}$ & 0 & $\frac{\sqrt{2}}{4}$ & 0 & $-\frac{\sqrt{9}}{4}$ & 0 & & \\
     & $|\mathbb{E}_{1}^{2}\rangle$ &  & & $\frac{\sqrt{10}}{4}$ & 0 & $-\frac{\sqrt{4}}{4}$ & 0 & $\frac{\sqrt{2}}{4}$ & 0 & & \\
    \hline
    \multirow{8}[8]{*}{$\frac{7}{2}$} & $|\mathbb{E}_{0}^{0}\rangle$ & & 0 & $\frac{\sqrt{7}}{8}$ & 0 & $\frac{\sqrt{35}}{8}$ & 0 & $\frac{\sqrt{21}}{8}$ & 0 & $\frac{1}{8}$ &  \\
    & $|\mathbb{E}_{0}^{1}\rangle$ & & 0 & $\frac{\sqrt{25}}{8}$ & 0 & $\frac{\sqrt{5}}{8}$ & 0 & $-\frac{\sqrt{27}}{8}$ & 0 & $-\frac{\sqrt{7}}{8}$ &  \\
    & $|\mathbb{E}_{0}^{2}\rangle$ & & 0 & $\frac{\sqrt{27}}{8}$ & 0 & $-\frac{\sqrt{15}}{8}$ & 0 & $\frac{1}{8}$ & 0 & $\frac{\sqrt{21}}{8}$ &  \\
    & $|\mathbb{E}_{0}^{3}\rangle$ & & 0 & $\frac{\sqrt{5}}{8}$ & 0 & $-\frac{\sqrt{9}}{8}$ & 0 & $\frac{\sqrt{15}}{8}$ & 0 & $-\frac{\sqrt{35}}{8}$ &  \\
     & $|\mathbb{E}_{1}^{0}\rangle$ & & $\frac{1}{8}$ & 0 & $\frac{\sqrt{21}}{8}$ & 0 & $\frac{\sqrt{35}}{8}$ & 0 & $\frac{\sqrt{7}}{8}$ & 0 &  \\
     & $|\mathbb{E}_{1}^{1}\rangle$ & & $\frac{\sqrt{7}}{8}$ & 0 & $\frac{\sqrt{27}}{8}$ & 0 & $-\frac{\sqrt{5}}{8}$ & 0 & $-\frac{\sqrt{25}}{8}$ & 0 &  \\
     & $|\mathbb{E}_{1}^{2}\rangle$ & & $\frac{\sqrt{21}}{8}$ & 0 & $\frac{1}{8}$ & 0 & $-\frac{\sqrt{15}}{8}$ & 0 & $\frac{\sqrt{27}}{8}$ & 0 &  \\
     & $|\mathbb{E}_{1}^{3}\rangle$ & & $\frac{\sqrt{35}}{8}$ & 0 & $-\frac{\sqrt{15}}{8}$ & 0 & $\frac{\sqrt{9}}{8}$ & 0 & $-\frac{\sqrt{5}}{8}$ & 0 &  \\
    \hline
    \multirow{10}[10]{*}{$\frac{9}{2}$} & $|\mathbb{E}_{0}^{0}\rangle$ & 0 & $\frac{\sqrt{9}}{16}$ & 0 & $\frac{\sqrt{84}}{16}$ & 0 & $\frac{\sqrt{126}}{16}$ & 0 & $\frac{\sqrt{36}}{16}$ & 0 & $\frac{1}{16}$  \\
    & $|\mathbb{E}_{0}^{1}\rangle$ & 0 & $\frac{\sqrt{49}}{16}$ & 0 & $\frac{\sqrt{84}}{16}$ & 0 & $-\frac{\sqrt{14}}{16}$ & 0 & $-\frac{\sqrt{100}}{16}$ & 0 & $-\frac{\sqrt{9}}{16}$  \\
    & $|\mathbb{E}_{0}^{2}\rangle$ & 0 & $\frac{\sqrt{100}}{16}$ & 0 & 0 & 0 & $-\frac{\sqrt{56}}{16}$ & 0 & $\frac{\sqrt{64}}{16}$ & 0 & $\frac{\sqrt{36}}{16}$  \\
    & $|\mathbb{E}_{0}^{3}\rangle$ & 0 & $\frac{\sqrt{84}}{16}$ & 0 & $-\frac{\sqrt{64}}{16}$ & 0 & $\frac{\sqrt{24}}{16}$ & 0 & 0 & 0 & $-\frac{84}{16}$  \\
    & $|\mathbb{E}_{0}^{4}\rangle$ & 0 & $\frac{\sqrt{14}}{16}$ & 0 & $-\frac{\sqrt{24}}{16}$ & 0 & $\frac{\sqrt{36}}{16}$ & 0 & $-\frac{\sqrt{56}}{16}$ & 0 & $\frac{\sqrt{126}}{16}$  \\

    & $|\mathbb{E}_{1}^{0}\rangle$&  $\frac{1}{16}$ & 0 & $\frac{\sqrt{36}}{16}$ & 0 & $\frac{\sqrt{126}}{16}$ & 0 & $\frac{\sqrt{84}}{16}$ & 0 & $\frac{\sqrt{9}}{16}$ & 0 \\
    & $|\mathbb{E}_{1}^{1}\rangle$&  $\frac{\sqrt{9}}{16}$ & 0 & $\frac{\sqrt{100}}{16}$ & 0 & $\frac{\sqrt{14}}{16}$ & 0 & $-\frac{\sqrt{84}}{16}$ & 0 & $-\frac{\sqrt{49}}{16}$ & 0 \\
    & $|\mathbb{E}_{1}^{2}\rangle$&  $\frac{\sqrt{36}}{16}$ & 0 & $\frac{\sqrt{64}}{16}$ & 0 & $-\frac{\sqrt{56}}{16}$ & 0 & 0 & 0 & $\frac{\sqrt{100}}{16}$ & 0 \\
    & $|\mathbb{E}_{1}^{3}\rangle$&  $\frac{\sqrt{84}}{16}$ & 0 & 0 & 0 & $-\frac{\sqrt{24}}{16}$ & 0 & $\frac{\sqrt{64}}{16}$ & 0 & $-\frac{\sqrt{84}}{16}$ & 0 \\
    & $|\mathbb{E}_{1}^{4}\rangle$&  $\frac{\sqrt{126}}{16}$ & 0 & $-\frac{\sqrt{56}}{16}$ & 0 & $\frac{\sqrt{36}}{16}$ & 0 & $-\frac{\sqrt{24}}{16}$ & 0 & $\frac{\sqrt{14}}{16}$ & 0 \\
    \hline
    \end{tabular}%
  \label{tab: spin cat-$x$ error basis}%
\end{table*}%

The spin cat-$x$ state is analogous to the bosonic (Schr\"{o}dinger) cat state, with codewords $|\mathcal{C}_\alpha^{\pm}\rangle \propto |\alpha\rangle \pm |-\alpha\rangle$, where $|\alpha\rangle$ is a coherent state of a single bosonic mode \cite{puri2020bias}.
The codewords of spin cat code, however, are orthogonal, which is not the case for the bosonic cat codewords. 
We also point out that the spin cat-$x$ code is exactly equivalent to the spin binomial code \cite{michael2016new,chiesa2020molecular, petiziol2021counteracting}, where
\begin{eqnarray}
|0_{\text{cat}}\rangle_x &=& |0_{\text{bin}}\rangle \equiv \frac{1}{\sqrt{2^{2I-1}}}\sum_{\substack{k=0\\k\in \text{even}}}^{2I}\sqrt{\binom{2I}{k}}|k-I\rangle\, , \nonumber \\
|1_{\text{cat}}\rangle_x &=& |1_{\text{bin}}\rangle \equiv \frac{1}{\sqrt{2^{2I-1}}}\sum_{\substack{k=0\\k\in \text{odd}}}^{2I}\sqrt{\binom{2I}{k}}|k-I\rangle\, .
\end{eqnarray}
For convenience we have also used the Holstein-Primakoff transformation here, such that we can interchangeably write the spin eigenstates in terms of Fock basis, i.e, $|I,m_I\rangle \to |n = I+m_I\rangle$. 

In \cite{grimsmo2020quantum}, it is shown that the bosonic cat and binomial code, among many other bosonic codes, belong to the large family of {\it rotationally symmetric bosonic (RSB)} codes --- codes that exhibit discrete rotation symmetry in the phase space.
Interestingly, one can also adapt the RSB code's construction for spin-$I$ systems, which we shall refer to as {\it rotationally symmetric spin (RSS)} codes. 
The basic formalism can be translated easily by taking the same Holstein-Primakoff transformation used throughout this work, namely, we take $\hat{n} \to I + \hat{I}_z$.
The $N$-order discrete rotation operator, the central tool defining the RSS code, then takes the form
\begin{equation}
\hat{R}_N = e^{i\frac{2\pi}{N}\left(I + \hat{I}_z\right)}\, .
\end{equation}
Moreover, the operator $\hat{Z}_N \equiv \hat{R}_{2N}$ acts as the logical Pauli $Z$ operator.
The logical codewords for any $N$-order RSS code can be constructed from discrete rotation superpositions of a normalized primitive state $|\Theta\rangle$.
In particular, it takes the form
\begin{equation}
|\mu_{N,\Theta}\rangle = \frac{1}{\sqrt{\mathcal{N}_\mu}} \sum_{n=0}^{2N-1}(-1)^{\mu m}e^{i\frac{m\pi}{N}\left(I + \hat{I}_z\right)}|\Theta\rangle\, ,\quad \mu=0,1\, , 
\end{equation}
where $\mathcal{N}_\mu$ are normalization constants.
As a result, we have the basic relation $\hat{Z}_N|\mu_{N,\Theta}\rangle = (-1)^\mu|\mu_{N,\Theta}\rangle$.
It is easy to see that the spin cat-$x$ code is an RSS code with $N=1$ and primitive state $|\Theta\rangle = |I,I\rangle_x$. 
More correspondence between RSS and RSB can be made but we leave this as future work.

\section{C. Error correction protocol}\label{sec: ideal EC protocol}

The error correction protocol for spin cat codes has been discussed before in Refs.~\cite{chiesa2020molecular, petiziol2021counteracting}.
Moreover, they have also provided the pulse sequence for the implementation of the error correction circuit. 
Meanwhile, Ref.~\cite{gross2024hardware} also provided a variation of the error correction protocol for the so-called MAUS (moment angular system) code, which is equivalent to the spin cat codes via an SU(2) rotation.
However, for completeness, we detail in here our variation of the error correction protocol utilizing error-transparent gates whenever it is appropriate.

\begin{figure*}[!htb]
\centering
\includegraphics[width=1\linewidth]{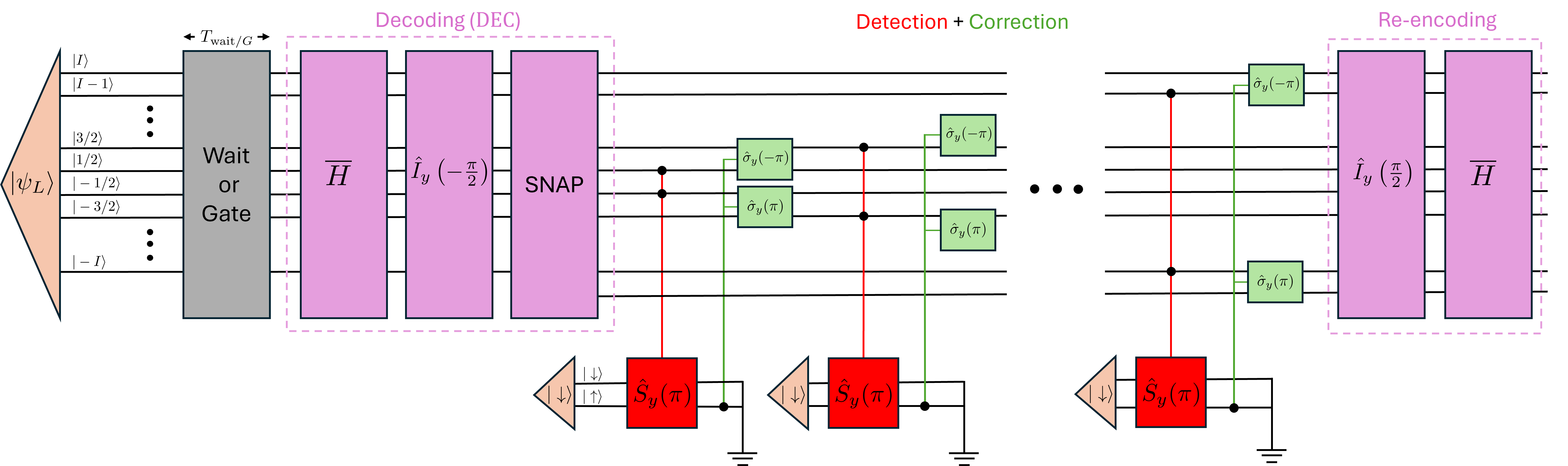}
\caption{\justifying{Illustration of the circuit to implement the error correction protocol for spin-$I$ cat-$x$ code. Each of the lines corresponds to the individual nuclear and electron state, with the labels above the line indicating the corresponding state. The initial state is allowed to evolve freely or under a gate Hamiltonian for some waiting/gate time $T_{\mathrm{wait}/G}$. The decoding step then prepares the encoded state such that logical and error information can be read out in the spin $z$ basis. An ancilla electron is used to detect and assist in correcting the encoded state, where it is optimal to reset the electron after each recovery step. After the recovery, the state can be re-encoded in the codespace.} }
\label{fig: ideal EC protocol general}
\end{figure*}

Since the spin cat-$x$ codes satisfy the Knill-Laflamme condition for the rank-$r$ dephasing noise, we can formulate the ideal detection and recovery procedure suggested by Knill and Laflamme \cite{knill1997theory}. 
The basic idea behind the error correction protocol consists of performing a projection on each error subspace $\mathcal{P}_k$, $k=1,\dots,r$, followed by performing correction for the errorwords to their corresponding codewords $|\mathbb{E}_{\mu}^k\rangle \to |\mu_L\rangle$.
We illustrate the error correction protocol in Fig. \ref{fig: ideal EC protocol general}, where we have designed it in a measurement-free error correction scheme \cite{pazsilva2010fault}.
In the following, we discuss the steps and ideas behind the error correction protocol.

Assume that the starting point is the logical state $\rho_L(0) = |\psi_L\rangle\langle\psi_L|$, $|\psi_L\rangle = \alpha |0_L\rangle + \beta|1_L\rangle$. 
We then let the state to dephase for some time resulting to $\rho_L(t) = \mathcal{N}(\rho_L(0)) = \sum_{k=0}E_k \rho_L(0) E_k^\dagger$, where $E_k = \sqrt{(\Gamma_n t)^k/k!}\hat{I}_z^k e^{-(\Gamma_n t/2)\hat{I}_z^2}$ is the dephasing error operator with (nuclear) dephasing rate $\Gamma_n = 2/T_2$, and $T_2$ is the dephasing time.
Thus, the action of the error channel after some waiting time $T$ can be approximated as
\begin{equation}
|\psi^{\text{err}}_L\rangle \approx \sqrt{1-\varepsilon}\left(\alpha |0_L\rangle + \beta|1_L\rangle \right) + \sqrt{\varepsilon_1}\left(\alpha |\mathbb{E}_0^1\rangle + \beta|\mathbb{E}_1^1\rangle \right) + \dots + \sqrt{\varepsilon_r}\left(\alpha |\mathbb{E}_0^r\rangle + \beta|\mathbb{E}_1^r\rangle \right) + \mathcal{O}(r+1)\, ,
\end{equation}
where $\varepsilon_k \sim (\Gamma_n T)^k$ is the probability of the $k$-th order error occurring. 
Moreover, since we are only able to detect and correct errors up to the $r$-th order, we leave the uncorrectable terms as $\mathcal{O}(r+1)$.
We now apply the decoding step $\hat{U}_{\text{dec}} = (\hat{U}_{\text{SNAP}})(e^{i\frac{\pi}{2}\hat{I}_y})\overline{H}$ and look at its effect on the corrupted state step-by-step. 
Firstly, applying $\overline{H}$ gives us
\begin{eqnarray}
\overline{H}|\psi^{\text{err}}_L\rangle & \approx & \sqrt{1-\varepsilon}\left(\alpha |\mathbb{D}_0^0\rangle  + \beta|\mathbb{D}_1^0\rangle \right) + \sqrt{\varepsilon_1}\left(\alpha |\mathbb{D}_0^1\rangle + \beta|\mathbb{D}_1^1\rangle \right) + \dots + \sqrt{\varepsilon_r}\left(\alpha |\mathbb{D}_0^r\rangle + \beta|\mathbb{D}_1^r\rangle \right) + \mathcal{O}(r+1) \nonumber \\
&=& \sqrt{1-\varepsilon}(\alpha |I,I\rangle_x + \beta|I,-I\rangle_x) - \sqrt{\varepsilon_1}(\alpha |I,I-1\rangle_x + \beta|I,-I+1\rangle_x) + \dots \nonumber \\
& & \qquad \qquad + \sqrt{\varepsilon_r}(\alpha |I,\frac{1}{2}\rangle_x + \beta|I,-\frac{1}{2}\rangle_x) + \mathcal{O}(r+1)\, , \nonumber 
\end{eqnarray}
where $|\mathbb{D}_{\mu}^{k}\rangle = \frac{1}{\sqrt{2}}\left(|\mathbb{E}_0^k\rangle +(-1)^\mu |\mathbb{E}^k_1\right) = (-1)^k|I,(-1)^\mu(I-k)\rangle$ is the dual of the errorwords. 
Then applying SU(2) rotation around the $y$-axis rotates the spin eigenstates from the $x$ basis to the $z$ basis, we have:
\begin{eqnarray}
e^{i\frac{\pi}{2}\hat{I}_y}\overline{H}|\psi^{\text{err}}_L\rangle &=& \sqrt{1-\varepsilon}(\alpha |I,I\rangle + \beta|I,-I\rangle) - \sqrt{\varepsilon_1}(\alpha |I,I-1\rangle + \beta|I,-I+1\rangle) + \dots \nonumber \\
& & \qquad \qquad + \sqrt{\varepsilon_r}(\alpha |I,\frac{1}{2}\rangle + \beta|I,-\frac{1}{2}\rangle) + \mathcal{O}(r+1) \, ,\nonumber 
\end{eqnarray}
Lastly, notice that there is a -1 phase accrued in each of the $k$-th errorwords with odd $k$, so we apply a SNAP gate to remove these phases to finally yield
\begin{eqnarray}
|\psi_L^{\text{err+dec}}\rangle &=& \hat{U}_{\text{dec}}|\psi^{\text{err}}_L\rangle \nonumber \\
&=& \sqrt{1-\varepsilon}(\alpha |I,I\rangle + \beta|I,-I\rangle) + \sqrt{\varepsilon_1}(\alpha |I,I-1\rangle + \beta|I,-I+1\rangle) + \dots + \sqrt{\varepsilon_r}(\alpha |I,\frac{1}{2}\rangle + \beta|I,-\frac{1}{2}\rangle) + \mathcal{O}(r+1)\, . \nonumber 
\end{eqnarray}
At this stage, we have turned the $k$-th errorwords to be detectable in the $\{|\pm(I-k)\rangle\}$ subspace of the spin eigenstates. 

To detect the different errors, we now introduce the electron ancilla initialized in the state $|\downarrow\rangle$. 
To detect the presence of $k$-th order error, we simply perform ESR $\pi$-pulse conditional on the spin states $|\pm(I-k)\rangle$. 
If an error is present, then the electron state will flip $|\downarrow\rangle \to |\uparrow\rangle$.
To perform correction, we then want to perform an NMR $\pi$-pulse on the $\vert\uparrow\rangle$ branch to flip the nuclear state $|\pm(I-k)\rangle \to |\pm(I-k+1)\rangle$. 
Note that this operation corresponds to the unitary $R^{(m_I)}_y(\theta) = \exp\left[-i\frac{\theta}{2}\hat{\sigma}^{(m_I)}_y\right]$, where $\hat{\sigma}^{(m_I)}_y$ is the Pauli-$y$ operator acting on the subspace spanned by $\{|m_I\rangle, |m_I-1\rangle\}$ with $\theta=\pm \pi$ depending on the nuclear spin state it wants to correct for. 
In an attempt to correct all errors, we successively perform a sequence of error detection and correction from the largest to smallest $k$ , thus consisting of $r-1$ recovery steps for spin-$I$ system. 
In between the successive recovery steps, we find that resetting the electron spin state is optimal if one considers a constant dephasing environment for the electron.
This is because the duration of the NMR $\pi$-pulse is around $0.5$ ms, which could have dephased the electron spin state considerably after multiple $\pi$-pulses.
For an error-free electron, resetting is not necessary. 
At the end of the recovery step, the state is maximally corrected modulo the $\mathcal{O}(r+1)$ terms, which correspond to the uncorrectable terms. 
After the recovery, the state is still encoded in the $z$-basis, which is prone to dephasing errors. 
Thus, we can re-encode it by performing the reverse operation of $\hat{U}_{\text{dec}}$ and ignoring the SNAP gate part since there is no phase correction required. 


\section{D. Fault-tolerance formalism for spin codes}\label{appendix: FT formalism}

In this section, we provide the definition of fault-tolerant gadgets of quantum computing based on the definition given in \cite{aliferis2006quantum, gottesman2009introduction}.
We adapt the definitions for the case of qudit spin codes with rank-$r$ errors -- analogous to multi-qubit codes capable of correcting errors up to $r$ weight.
This generalization is similar to the generalization done in the bosonic code literature \cite{xu2024fault, my2025circuit}.

\begin{definition}[Fault-Tolerant Measurement (Spin Code)]\label{def: FT measurement}
A measurement gadget is said to be fault-tolerant against rank-$r$ faults if there are rank-$s$ errors in the input and there are rank-$t$ errors in the measurement gadget itself, such that $s+t\leq r$. Moreover, the same result out of the measurement gadget should be equal as if we had performed ideal decoding on the incoming state and ideally measuring the qubit.
\end{definition}

\begin{definition}[Fault-Tolerant Preparation (Spin Code)]\label{def: FT preparation}
A preparation gadget is said to be fault-tolerant against rank-$r$ faults if the preparation step output a state that is within $s\leq r$ errors of a properly encoded state. Moreover, after ideal decoding the state be equal to the ideal target state.
\end{definition}

\begin{definition}[Fault-Tolerant Gate (Spin Code)]\label{def: FT gate}
A gate gadget is said to be fault-tolerant against rank-$r$ faults if there are rank-$s$ errors in the input and at most $t$ faults occur in the gate protocol, such that $s+t \leq r$. Moreover, the same result out of the gate gadgets should be equal as if we had performed ideal decoder on the incoming state then the ideal gate is acted on it.
\end{definition}

\begin{definition}[Fault-Tolerant Error Correction (Spin code)]\label{def: FT EC}
An error correction (EC) gadget is said to be fault-tolerant against rank-$r$ faults if the error correction step output a state that it within $t \leq r$ errors of a properly encoded state. Moreover, if the incoming state has rank-$s$ errors and there are rank-$t$ errors in the EC gadget, such that $s+t\leq r$, then ideally decoding the output state should be the same as ideally decoding the input state.
\end{definition}

The ideal decoding, here, refers to the performing ideal error correction and decoding the logical state into qubit state. 
Note that since we are primarily focused on dephasing errors, all the definitions above and upcoming discussions are assumed with errors from the set $\mathcal{E}_z^{[r]} = \{\one, \hat{I}_z,\hat{I}_z^2,\dots,\hat{I}_z^r\}$, but in principle can be extended easily to general errors of rank-$r$.
In that case, the error basis can be expressed as $\mathcal{E}^{[r]} = \{\hat{I}^{a}_x\hat{I}^b_y\hat{I}^c_z\, \vert \, 0 \leq a+b+c \leq r\}$.
For general errors, it is convenient to express the errors in terms of the irreducible spherical tensor operators $T^{(k)}_q(I)$. 
Interested readers are referred to \cite{omanakuttan2024fault,biedenharn1984angular}.

It is clear that the error-transparent gates discussed in the main text satisfy the definition of the FT gate following Definition \ref{def: FT gate}. 
In the following, we show that the measurement and Knill-EC circuit constructed in error-transparent manner from the main text indeed satisfy the FT definitions laid in the above.

\subsection{D.1 Fault-tolerance of ET measurement}

Here, we show that the ET measurement discussed in Results is FT following Definition \ref{def: FT measurement}. 
Consider an initial logical state $\rho_L = |\psi_L\rangle \langle \psi_L|$, $|\psi_L\rangle = \alpha|\overline{0}\rangle + \beta |\overline{1}\rangle$.
Assuming that the state experiences dephasing noise of order $s$, we can approximate the resulting state as statistical mixture of logical states transformed by increasing order of $\hat{I}_z$ up to $s$. 
This implies a noisy stochastic model that allows the state to be expressed as
\begin{equation}
\rho_L^{s-\text{err}} \approx (1-\epsilon)\rho_L + \epsilon_1 \rho_L^{\text{err}(1)} + \epsilon_2 \rho_L^{\text{err}(2)} + \dots + \epsilon_s \rho_L^{\text{err}(s)}\, ,
\end{equation}
where $\rho_L^{\text{err}(k)} = |\psi_L^{\text{err(k)}}\rangle \langle \psi_L^{\text{err(k)}}|$, $|\psi_L^{\text{err(k)}}\rangle = \alpha|\hat{I}_z^k\overline{0}\rangle + \beta |\hat{I}_z^k\overline{1}\rangle$, $ k=1,\dots,s$, with $|\hat{I}_z\overline{\mu}\rangle$ being the normalized state of $\hat{I}_z|\overline{\mu}\rangle$, and $\epsilon = \epsilon_1 + \dots + \epsilon_s$, $\epsilon_k\ll 1$.
The probabilities of the measurement outcome between the noiseless state $\rho_L$ and $s$-noisy state $\rho_L^{s-\text{err}}$ is the same, i.e.,
\begin{eqnarray}
\text{Pr}\left(z|\mathcal{M}_{\overline{Z}}^{\text{ET}}\right) &=& \tr{\rho_L \hat{M}_z^{Z,\text{ET}}} \nonumber \\ 
&=& \tr{\rho_L^{s-\text{err}} \hat{M}_z^{Z,\text{ET}}}\, ,\quad z=0,1
\end{eqnarray}
for $s \leq r$.
Moreover, if the measurement is noisy up to $t$-order, the measurement outcome probabilities are still exact for $s+t \leq r$.
This outcome is equivalent to ideally correcting the input state and performing ideal measurement.
Hence, it satisfies the FT requirement for measurement given in Definition \ref{def: FT measurement}.

\subsection{D.2 Fault-tolerance of Knill-EC}

\begin{figure}[!htb]
\centering
\includegraphics[width=0.5\linewidth]{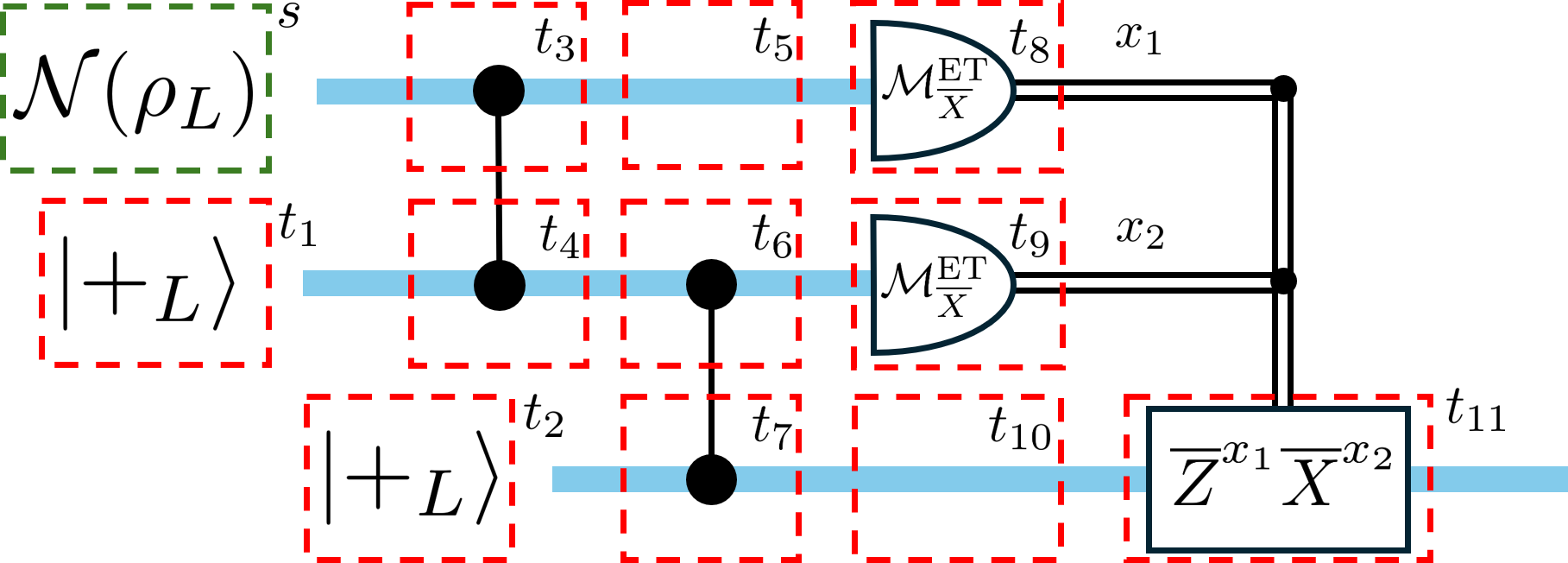}
\caption{\justifying{Faults and fault locations in Knill-EC. We assume the input state to have $s$-order noise, and the $j$-th components in the Knill-EC circuits with noise of order $t_j$.}}
\label{fig: knill ec fault locations}
\end{figure}

Here, we show that the Knill-EC circuit constructed using ET operations (Fig. \ref{fig: telecorrection circuit}) is FT following Definition \ref{def: FT EC}. 
The proof follows closely the steps laid out in Ref. \cite{my2025circuit}.
Assume that the input location has fault $s$. 
The Knill-EC gadget composed of 11 locations, denoted by $j\in \{1,2,\dots,11\}$, with faults $t_j$ (see Fig. \ref{fig: knill ec fault locations}). 
Following the dephasing noise model described in Results, we say that fault $t_j$ occurs at location $j$ if an error of the form $\hat{I}_z^{t_j}$ has occurred. 
To simplify the discussion, we model the faults allocation in the different circuit components as follows: the noisy preparation is equal to ideal preparation followed by noise, noisy gate is equal to noise preceding the ideal gate, and noisy measurement is equal to noise preceding the ideal measurement.

To understand the degree of faults encountered in each rail, we can propagate all the faults occurring in each rail to the end just before the ideal measurement.
Since all the operations are error-transparent with respect to the spin cat-$x$ codes initialized in each rail, the faults propagate directly to the end without any changes. 
In particular, the number of faults at the end of each rail is given by
\begin{eqnarray}
t^{(I)} &=& s + t_3 + t_5 + t_8\, , \\
t^{(A)} &=& t_1 + t_4 + t_6 + t_9\, , \\
t^{(O)} &=& t_2 + t_7 + t_{10} + t_{11}\, ,
\end{eqnarray}
where the superscripts $(I),(A),(O)$ correspond to the input, ancilla, and output rail, respectively. 
The total number of faults in the Knill-EC gadget is given by $t = \sum_{j=1}^{11}t_i$. 
Assuming $s + t \leq r$, one can immediately infer that $t^{(n)} \leq r$ for $n\in \{I,A,O\}$. 
This tells us that the measurement outcome of the ET $\overline{X}$ gadget is equivalent to the ideal case as demonstrated in the previous section. 
Therefore, the faults in the output state amounted to only $t^{(O)}$ and is independent of the fault occurred in the input state.
Since $t^{(O)} \leq r$, then it is clear that ideally decoding the output state is equal to ideally decoding the input state.
This shows that the constructed Knill-EC gadget satisfies the FT requirement for EC gadget given in Definition \ref{def: FT EC}. 


\section{E. Simulation details for Knill-EC}\label{appendix: Knill-EC numeric details}

In this section, we provide details on the different realizations of the logical dual measurement set $\{\texttt{DUAL},\texttt{CAN},\texttt{PG},\texttt{ET}\}$ and the numerical simulation discussed in Results.
The \texttt{DUAL} measurement corresponds to the {\it basic} logical dual measurement with measurement operator $\hat{M}^{X,\texttt{DUAL}}_0 = |+_L\rangle \langle +_L|$, $\hat{M}^{X,\texttt{DUAL}}_1 = |-_L\rangle \langle -_L|$.
Note that these operators are complete in the codespace, i.e., $\sum_{x=0}^1 \hat{M}^{X,\texttt{DUAL}}_x = \hat{P}_{\mathcal{C}}$. 
Hence, to satisfy the completeness relation one can add another POVM element $\one - \hat{P}_{\mathcal{C}}$. 

The \texttt{CAN} measurement corresponds to the {\it canonical phase} measurement discussed in \cite{grimsmo2020quantum}, where the construction there can be easily adapted for the spin cat-$x$ code since it belongs to the family of rotationally symmetric spin codes as discussed above.
The canonical phase measurement, introduced by Holevo \cite{holevo2011probabilistic} and Helstrom
\cite{helstrom1969quantum}, is an optimal measurement setting for determining the phase in the phase estimation problem $e^{i\theta\hat{n}}|\psi_0\rangle$, where $\hat{n}$ is the bosonic number operator and $|\psi_0\rangle$ is some fiducial state. 
In the spin case, we take the Holstein-Primakoff transformation $\hat{n} \to I + \hat{I}_z$. Let $|\psi_0\rangle = \sum_{n=0}^{2I}c_n|n\rangle$, the POVM operator for canonical phase measurement takes the form
\begin{equation}
\hat{M}^{\texttt{CAN}}(\theta) = \frac{1}{2\pi}\sum_{n,n' = 0}^{2I} \gamma_{n}\gamma_{n'}^* |n\rangle\langle n'|\, ,
\end{equation}
where $\gamma_n = c_n/|c_n|$ for $c_n\neq 0$ and we set $\gamma_n = 1$ otherwise. 
Since the estimation problem is to distinguish between $|\pm_L\rangle$, the fiducial state is set to be $|\psi_0\rangle = |+_L\rangle$ and we can grouped the POVM into dichotomic measurement $\hat{M}^{X,\texttt{CAN}}_{x=0/1} = \int_{\theta\in\mathcal{D}_{N=1,\pm}}d\theta\, \hat{M}^{\texttt{CAN}}(\theta)$, where $\mathcal{D}_{N,\pm}\subset [0,2\pi)$ is the domain of $\theta$ corresponding to identifying the state as $|+_L\rangle$ or $|-_L\rangle$. 
For $N$-order RSS code, this is given by
\begin{equation}
\mathcal{D}_{N,\pm} = \begin{cases}
\bigcup_{k=0}^{N-1} \left[(2k-\frac{1}{2})\frac{\pi}{N}, (2k+\frac{1}{2})\frac{\pi}{N})\right)\, ,\quad \text{for +} \\
\bigcup_{k=0}^{N-1} \left[(2k+\frac{1}{2})\frac{\pi}{N}, (2k+\frac{3}{2})\frac{\pi}{N})\right)\, ,\quad \text{for -}
\end{cases}
\end{equation}
As an example, for $N=2$, we have $\mathcal{D}_{2,+} = [\frac{-\pi}{4},\frac{\pi}{4}) \cup [\frac{3\pi}{4},\frac{5\pi}{4})$ and $\mathcal{D}_{2,-} = [\frac{\pi}{4},\frac{3\pi}{4}) \cup [\frac{5\pi}{4},\frac{7\pi}{4})$.
Note that we have the completeness relation $\sum_{x=0}^1 \hat{M}^{X,\texttt{CAN}}_{x} = \one$.
In contrast to the \texttt{DUAL} measurement, the \texttt{CAN} measurement cannot deterministically distinguish the logical codewords due to embedded phase uncertainty. 

Lastly, we have the {\it pretty good} (\texttt{PG}) measurement \cite{hausladen1994prettygood} that is considered to be near-optimal in discriminating states.
The POVM operator takes the form
\begin{equation}
\hat{M}^{X,\texttt{PG}}_{x=0/1} = \hat{\sigma}^{-\frac{1}{2}}\mathcal{N}\left(|\pm_L\rangle \langle \pm_L|\right)\hat{\sigma}^{-\frac{1}{2}}\, ,
\end{equation}
where $\hat{\sigma} = \mathcal{N}(\hat{P}_\mathcal{C})$. 
Similar to the \texttt{DUAL} measurement, the \texttt{PG} measurement operators are complete in the codespace, so one can add a POVM element $\one - \hat{P}_{\mathcal{C}}$ to satisfy the completeness relation.
The \texttt{PG} measurement is designed with the knowledge of the noise profile $\mathcal{N}$ as a means to counteract the effect of noise in the incoming state to be measured.

Unlike the simulation for the error correction (Figs. \ref{fig: ET X gate fidelities}-\ref{fig: ET $CZ$ gate performance complete}), here we simulate the Knill-EC gadget by assuming that it is noiseless and independent of the physical parameters. 
As a result, the output of the state after correction has an analytical solution \cite{hillmann2022performance}:
\begin{equation}
\mathcal{R}_{\text{Knill}}^{\texttt{T}} \circ \mathcal{N}(\rho_L) = \frac{1}{4}\sum_{i,j=0}^3c^{\texttt{T}}_{ij}(\Vec{x})P^{\dagger}_{i^*(\Vec{x})}P_i\rho_L P_j^\dagger P_{i^*(\Vec{x})}\, ,
\end{equation}
where the weights $c^{\texttt{T}}_{ij}(\Vec{x}) = \tr{\hat{M}^{X,\texttt{T}}_{x_1} \otimes\hat{M}^{X,\texttt{DUAL}}_{x_2}\hat{\sigma}_{ij}}$ are associated with measurement outcomes $\Vec{x} = (x_1,x_2)$. 
As discussed in the main text, we only vary the different measurement realizations in the input rail, and fix the measurement in the ancilla rail to be \texttt{DUAL}.
The operators $P_i\in \{P_L,Z_L,X_L,X_LZ_L\}$ represents logical Pauli operators encoded in the spin cat-$x$ code. 
The operator $\hat{\sigma}_{ij}$ represents the damaged two-mode dual-basis code states that are obtained by commuting the noise channel through the $\overline{CZ}$ gate, that is
\begin{equation}
    \hat{\sigma}_{ij} = \mathcal{U}_{CZ} \circ \mathcal{N} \circ \mathcal{U}_{CZ}^{\dagger} (|i\rangle\langle j|)\, ,
\end{equation}
where $|i\rangle,|j\rangle \in \{|+_L\rangle, |-_L\rangle\}$.
To recover the original logical state, the most likely Pauli correction $P_{i^*(\Vec{x})}$ is used based on the maximum likelihood decoder
\begin{equation}
i^*(\Vec{x}) = \max_{i=0,1,2,3} \tr{\hat{M}^{X,\texttt{T}}_{x_1}\otimes \hat{M}^{X,\texttt{DUAL}}_{x_2}\hat{\sigma}_{ii}}\, .
\end{equation}

\end{appendix}

\end{document}